\documentclass[twocolumn, apj, numberedappendix, iop]{openjournal}
\usepackage{natbib}
\usepackage{graphicx,amsmath,amssymb,amstext}
\usepackage{amsbsy,amsfonts,amsthm,color}

\usepackage[colorlinks,linkcolor=blue,citecolor=blue,urlcolor=blue ]{hyperref}
\usepackage[utf8]{inputenc}
\usepackage{float}
\usepackage{graphicx}
\usepackage{orcidlink}
\usepackage{booktabs}
\usepackage{xspace}

\usepackage{xcolor}
\definecolor{darkgreen}{rgb}{0.0,0.45,0.0}

\newcommand{\galacticus}{\textsc{Galacticus}\xspace}

\def\msun{\mathrm{\, M_\odot}}
\def\kms{\mathrm{\, km \, s^{-1}}}

\defcitealias{2012ApJ...744..159L}{L12}
\newcommand{\ltw}{\citetalias{2012ApJ...744..159L}\xspace}

\begin{document}

\title{Accelerated calibration of semi-analytic galaxy formation models\vspace{-4ex}}

\author{Andrew Robertson$^{*}$\orcidlink{0000-0002-0086-0524}}
\author{Andrew Benson\orcidlink{0000-0001-5501-6008}}
\email{$^*$email: arobertson@carnegiescience.edu}
\affiliation{The Observatories of the Carnegie Institution for Science, 813 Santa Barbara Street, Pasadena, 91101, CA, USA}

\begin{abstract}
We present an accelerated calibration framework for semi-analytic galaxy formation models, demonstrated with \galacticus. Rather than fitting directly to properties such as the low-redshift stellar mass function (SMF)---which requires evolving thousands of halos per likelihood evaluation---we construct a fast likelihood from the stellar-to-halo mass relation (SHMR; mean and scatter) evaluated at a small set of target halo masses, reducing each evaluation to simulating only tens of galaxies. We sample the posterior over \galacticus\ parameters with Markov Chain Monte Carlo and show that the resulting calibration reproduces the low-redshift SMF. We then extend the method to additional datasets, using a higher-redshift SHMR and the low-redshift stellar mass--size relation as examples, and assess performance for large scale structure survey-relevant properties: stellar masses, sizes, and emission-line strengths. The SMF matches data well at low redshift, but toward higher redshift the model yields too few massive galaxies and too many low-mass galaxies. Size evolution with redshift is approximately correct, but the mass--size relation is too flat, producing massive galaxies that are too small. The H$\alpha$ luminosity function is well reproduced at $z\sim2$, but by $z\sim0.4$ the model overproduces highly star-forming, H$\alpha$-bright systems. These discrepancies suggest the model lacks sufficient flexibility (e.g. in gas cooling/recycling or feedback) to reconcile all datasets simultaneously. Our strategy complements emulator-based methods for calibrating semi-analytic models by enabling rapid, low-cost scans of model choices and parameterisations---a capability we envision leveraging to supply calibrated starting points for more detailed follow-up inference.
\end{abstract}

\keywords{}

\maketitle

\section{Introduction}

Semi-analytic models (SAMs) of galaxy formation have proven to be powerful tools for interpreting large-scale structure (LSS) surveys. By combining physically motivated prescriptions for baryonic processes (e.g. gas cooling, star formation, feedback from stars and supermassive black holes) with halo merger trees from dark matter simulations or extended Press-Schechter theory, SAMs enable predictions for a wide range of galaxy properties across cosmic time \citep[e.g.][see \citealt{2015ARA&A..53...51S} for a review]{1991ApJ...379...52W, 1999MNRAS.303..188K, 2000MNRAS.319..168C, 2006MNRAS.368.1540C, 2006RPPh...69.3101B, 2008MNRAS.391..481S, 2016ApJS..222...22C, 2017MNRAS.471.1401C, 2018MNRAS.481.3573L, 2020MNRAS.491.5795H, 2024A&A...687A..68D, 2024PASA...41...53S, 2024MNRAS.531.3551L}. These models strike a balance between computational speed and physical complexity, enabling rapid exploration of model predictions. With the advent of upcoming LSS surveys from Euclid \citep{2025A&A...697A...1E}, the Nancy Grace Roman Space Telescope \citep{2015arXiv150303757S, 2021MNRAS.507.1746E}, and the Vera C. Rubin Observatory's Legacy Survey of Space and Time \citep{2019ApJ...873..111I}, we will have access to galaxy datasets of unprecedented volume and precision. These surveys will place stringent constraints on models of galaxy formation, and SAMs are ideally suited for generating mock catalogues, testing for sources of systematic error, and interpreting observed trends in a physically motivated framework.

In order to produce realistic populations of galaxies, SAMs must be carefully calibrated against observational data. This is necessary because many of the baryonic processes that drive galaxy evolution are complex and poorly constrained from first principles. Consequently, SAMs include a number of free parameters that must be adjusted to match observations. Early efforts to calibrate SAMs relied heavily on ``hand tuning'' of parameters \citep[e.g.][]{2000MNRAS.319..168C, 2006MNRAS.370..645B}, where modellers -- guided by physical intuition -- adjusted parameters iteratively based on visual comparisons of the model predictions with observations such as luminosity functions or stellar mass functions. This method can be effective when targeting a single observable \citep[e.g., the $z \sim 0$ stellar mass function in][]{2016ApJS..222...22C}, particularly if the number of parameters varied is small. However, galaxy formation physics is highly non-linear, and different observables are often sensitive to overlapping subsets of parameters \citep[e.g. stellar feedback affects both stellar masses and sizes of galaxies; ][]{2010MNRAS.407.2017B, 2016MNRAS.462.3854L}. As such, manual calibration becomes increasingly difficult as the number of observables increases or when aiming to capture more subtle correlations and trends in the data.

To address the shortcomings of manual calibration, various more-automated means of exploring the parameter-space of SAMs have been used. Markov Chain Monte Carlo (MCMC) is one such approach, which produces draws from the posterior distribution of model parameters \citep[for a review of MCMC as applied to problems in astrophysics, see][]{2017ARA&A..55..213S}. We note that the goal of MCMC algorithms is not to find models that fit the data especially well, rather it is to generate draws from the posterior distribution for the model parameters, that will typically cluster around the best-fitting model in something known as the \emph{typical set} \citep{2017arXiv170102434B}. This means that MCMC is particularly useful when we are interested not only in finding a good model, but in understanding how the various SAM parameters interact with one another, and therefore what alternative models can also provide a good fit to the data. That said, the burn-in stage of many MCMC methods is also a reasonably efficient and reliable way to find parameters close to those that are optimal.

Early attempts at using MCMC to calibrate SAMs include \citet{2008MNRAS.384.1414K}, \citet{2009MNRAS.396..535H} and \citet{2011MNRAS.416.1949L}. These demonstrated that -- when fitting to just the stellar mass function at low redshift (or similarly, to a low-redshift K-band luminosity function) -- the posterior distribution is highly non-Gaussian, with complex degeneracies between different model parameters. Subsequent work extended these approaches to include multiple observational constraints across redshift (such as stellar mass functions, gas fractions, and quiescent fractions) to help reduce degeneracies and improve physical realism \citep[e.g.][]{2013MNRAS.431.3373H, 2015MNRAS.451.2663H}. 

In this paper, we seek to calibrate the \galacticus SAM \citep{2012NewA...17..175B} in order to better reproduce key properties of observed galaxies. Compared to some earlier SAMs that were calibrated using MCMC, \galacticus predictions are substantially slower to evaluate. This is because \galacticus incorporates a detailed treatment of subhalo evolution \citep{2024PhRvD.110b3019D}, realistic models of star formation (which require the evaluation of integrals over the disk surface density profile; \citealt{2006ApJ...650..933B}), and detailed calculations of the sizes of galaxies and the adiabatic contraction of the dark matter halo (following the approach of \citealt{2000MNRAS.319..168C}). 

These more realistic physical ingredients necessarily increase the computation time. For example, the likelihood evaluations in \citet{2011MNRAS.416.1949L} contain 3000 merger trees, with final halos with masses ranging between $10^{10}$ and $10^{15} \msun$. They could evolve these 3000 trees with their SAM in approximately 1 minute, which -- combined with a parallel MCMC algorithm -- enabled them to perform of order a million likelihood evaluations during their MCMC. 
Their exact halo mass distribution is not stated, but assuming that their 2000 halos with masses above $10^{12.5} \msun$ were distributed as random draws from the halo mass function \citep[as found to be approximately optimal for minimal ``error per CPU hour'' on the stellar mass function in][]{2014MNRAS.444.2599B}, then this comparable calculation with \galacticus takes approximately 100 CPU hours. 

If we wanted to use MCMC in a similar manner as \citet{2011MNRAS.416.1949L}, then our $\sim$ four orders of magnitude slower-to-run model would mean a one-million-sample MCMC would take approximately 100 million CPU hours. This is relatively infeasible even as a one-off, and makes using this method of calibration prohibitively expensive given that one often wishes to perform the calibration multiple times (fitting to different target datasets, using different models, varying different parameters, etc.).  

Because the computation time is the product of the number of likelihoods that need to be evaluated, and the evaluation time of a single likelihood, there are two obvious strategies for trying to speed up calibration. The first is to reduce the number of likelihood evaluations required, the second is to speed up the likelihood evaluations. Adopting the first strategy, \citet{2015ApJ...801..139R} found around a factor of ten speed up by using particle swarm optimisation \citep{488968} instead of MCMC, and this method has since been used to calibrate other SAMs \citep{2024PASA...41...53S, 2024MNRAS.531.3551L}.

Another approach to decreasing the number of full likelihood evaluations required to explore the parameter space of a SAM is to use a relatively small number of full model runs as training data for a surrogate model. This approach was first demonstrated in \citet{2010MNRAS.407.2017B} for the GALFORM SAM, where a technique known as \emph{Bayesian history matching} \citep{2014arXiv1405.4976V} was used. The basic idea was to build an emulator for the output of GALFORM, that could predict the expected value (with an associated uncertainty) for various observed quantities as a function of the GALFORM parameters. Regions of parameter space that the emulator was confident led to a poor match to observations were discarded, while additional GALFORM models were run in the remaining regions, focusing on places where the model was potentially good but the emulator was uncertain. Repeating this, the region of parameter space containing the models best able to explain the observations was uncovered.

Recent work along similar lines used deep learning to emulate the predictions of GALFORM \citep{2021MNRAS.506.4011E}. More specifically, they trained a neural network to predict population-level statistics of GALFORM, such as the luminosity function in various photometric bands, the black hole mass--bulge mass relation, etc. They trained the neural network with around one thousand runs of GALFORM and then demonstrated that its predictions were a reasonable match to the truth for additional GALFORM parameter combinations not used in the training data. The neural network predictions are fast to generate, and so once the emulator has been trained, an MCMC can be efficiently run using the neural network for the likelihood evaluations.

While this emulator-based approach reduces the number of costly full-model evaluations, another strategy is to reduce the cost of each evaluation, allowing for a more direct and flexible exploration of parameter space. This is the approach we investigate in this paper. The key insight we build on is that while one cannot directly control the stellar masses, luminosities, or sizes of galaxies produced by a semi-analytic model, one \emph{can} choose which halo masses to simulate. This opens up a natural route to accelerate model calibration: by focusing on only a small number of halo mass bins and asking whether the model produces galaxies in those halos that match specific observational constraints.

For instance, considering the stellar masses of galaxies, we can attempt to match the observed stellar-to-halo mass relation (SHMR). One might expect that if a model accurately reproduces the galaxy stellar mass distribution in halos of mass $10^{11}$ and $10^{12}\msun$, then it may also perform well for intermediate masses such as $10^{11.5} \msun$, or perhaps even for slightly extrapolated regimes such as $10^{12.5} \msun$. Testing this idea in the context of \galacticus is the central focus of this paper. We demonstrate that calibrating the model to match the SHMR at a few well-chosen halo mass scales provides an efficient route to recovering a good match to the observed stellar mass function. We then show how this approach can be naturally extended to other galaxy properties, using galaxy sizes as a concrete example.

This paper is structured as follows. In Section~\ref{sect:galacticus} we provide an overview of the \galacticus model, with special emphasis on the prescriptions within the model that we vary during our calibration process. In Section~\ref{sect:calibration} we describe our calibration strategy, starting with a fit to the SHMR at a single redshift, and then extending this to consider the SHMR at two redshifts as well as constraints on the sizes of galaxies. In Section~\ref{sect:results} we take our calibrated model and compare it against observational data of the galaxy stellar mass function, H$\alpha$ emission line luminosity function, and stellar mass--size relation, each over  the redshift range $0 \lesssim z \lesssim 2.5$. In Section~\ref{sect:discussion} we discuss potential avenues to improve our method, and what additional freedom we may need to give \galacticus to simultaneously match galaxy sizes and stellar masses across cosmic time. Finally we present our conclusions in Section~\ref{sect:conclusions}.

\section{The Galacticus semi-analytic Model}
\label{sect:galacticus}

\galacticus is a flexible and modular semi-analytic model of galaxy formation and evolution \citep{2012NewA...17..175B}. It was designed from the outset to incorporate a wide range of physical processes, and to allow users to readily explore alternative prescriptions for galaxy formation physics. 
As previously noted, \galacticus is considerably slower to run than some previous SAMs \citep[e.g.][]{2009MNRAS.396..535H, 2011MNRAS.416.1949L}, which is the main driver for having to find alternative calibration strategies to ``brute force'' MCMC. \galacticus incorporates a much more detailed treatment of subhalo evolution\footnote{Which avoids problems associated with artificial disruption of subhalos \citep{2021MNRAS.505...18E}, such as orphan galaxies \citep{2022MNRAS.510.2900D}.} \citep{2024PhRvD.110b3019D}, tracking tidal mass loss and heating through a multi-level hierarchy (i.e., including sub-subhalos, etc.), more realistic models of star formation (which require the evaluation of integrals over the disk surface density profile; \citealt{2006ApJ...650..933B}), and detailed calculations of the sizes of galaxies and the adiabatic contraction of the dark matter halo (following the approach of \citealt{2000MNRAS.319..168C}). The simpler models used by \citet{2009MNRAS.396..535H} and \citet{2011MNRAS.416.1949L} allowed the evolution of some properties across timesteps (e.g. the mass of stars formed) to be computed analytically (with some further simplifying assumptions, such as that the structure of the galaxy does not evolve across the timestep), following approaches similar to those described in detail by \cite{2000MNRAS.319..168C}.

We note that several modern SAMs have incorporated increasingly sophisticated physical prescriptions over the past decade. For example, models such as GAEA \citep{2024A&A...687A..68D}, Shark \citep{2024MNRAS.531.3551L}, and more recent versions of \textsc{L-Galaxies} \citep{2020MNRAS.491.5795H} and the Santa Cruz SAM \citep{2015MNRAS.453.4337S} include H$_2$-regulated star formation, detailed treatments of environmental and black-hole driven feedback, and in the case of \textsc{L-Galaxies}, a comprehensive MCMC prescription. As such, the contrast drawn here with the earlier SAMs used in \citet{2009MNRAS.396..535H} and \citet{2011MNRAS.416.1949L} should be understood as a comparison to the specific models employed in those early MCMC studies, rather than to the current state of semi-analytic modelling more broadly.

Furthermore, while many modern SAMs now employ some form of sub-timestep evolution, \galacticus performs a fully coupled ODE integration with adaptive, accuracy-controlled timesteps, which remains relatively uncommon and is a major contributor to its higher computational cost. In the MCMC studies of \citet{2009MNRAS.396..535H} and \citet{2011MNRAS.416.1949L}, the evolution of galaxy properties was effectively advanced with an explicit, single-step scheme on a fixed time grid tied to the $N$-body simulation outputs, subdivided into a modest number of sub-steps. For example, the \citet{2009MNRAS.396..535H} implementation in the \textsc{L-Galaxies} framework uses timesteps of order $10$–$20$~Myr. By contrast, \galacticus employs the Cash–Karp \citep{10.1145/79505.79507} method to solve the coupled set of differential equations that describe the evolution of each galaxy, with adaptive timesteps chosen to ensure a specified accuracy per step.\footnote{A similar ODE-integration scheme with adaptive timesteps is used in \textsc{Shark} \citep{2018MNRAS.481.3573L}.} For the accuracy choice we employed in this work (0.01\% accuracy in each step), our final calibrated model (see Section ~\ref{sect:results_fullCalibration}) has median timesteps of around $1$~Myr, with the shortest timesteps being a couple of orders of magnitude shorter. These factors -- in particular the fully coupled, accuracy-controlled ODE integration together with the detailed subhalo hierarchy -- result in \galacticus being substantially slower than the SAM configurations used in the earlier MCMC studies.


In this section, we describe the key features of the \galacticus model that are most relevant to our paper, noting that our model choices are typically the same as those used in the \galacticus branch of the \textsc{MultiDark-Galaxies} project \citep{2018MNRAS.474.5206K}, as well as for generating mock Roman data for the High Latitude Spectroscopic Survey \citep{2022ApJ...928....1W}. Given that the \galacticus source code is publicly available,\footnote{\href{https://github.com/galacticusorg/galacticus/wiki}{github.com/galacticusorg/galacticus}} and that our configuration files are also available,\footnote{\href{https://github.com/Andrew-Robertson/CalibratingGalacticus-Paper1}{github.com/Andrew-Robertson/CalibratingGalacticus-Paper1} and \href{https://doi.org/10.5281/zenodo.16952803}{doi.org/10.5281/zenodo.16952803}.} we do not list the tens of model choices and hundreds of parameter values here, but instead summarise the key physics in \galacticus, with special emphasis on the definitions of parameters that we vary during our MCMCs.


\subsection{Merger trees}

In this work we use merger trees generated using the extended Press-Schechter (EPS) formalism, following the method of \citet{2008MNRAS.383..557P}. EPS-based trees have been extensively tested and shown to reproduce the primary statistical properties of halo merger histories found in cosmological simulations \citep{2014MNRAS.440..193J}. Their primary advantage for the present study is the high degree of control they offer over the halo population: final halo masses, redshift sampling, and mass resolution can be specified explicitly, and trees can be generated in arbitrary numbers at negligible computational cost. This is particularly well suited to our calibration strategy, which targets specific halo masses.

Merger trees extracted from $N$-body simulations have complementary strengths, most notably capturing halo assembly in a fully non-linear context and preserving spatial correlations between halos. However, such trees are tied to a fixed simulation volume and mass resolution, and provide only a discrete sampling of final halo masses, making it less straightforward to target specific masses. EPS-based trees, by contrast, lack spatial information and may not capture the full diversity of correlated assembly histories at fixed halo mass, but for our purposes -- where we focus on galaxy properties rather than spatial clustering -- their flexibility and efficiency outweigh these limitations.

\subsection{Gas cooling}

Cooling rates from the hot halo are computed using the approach introduced in \citet{2010MNRAS.405.1573B}. In this model, the radius at which the thermal energy of the gas is equal to the integral of the cooling rate over time is known as $r_\mathrm{cool}$, and gas within $r_\mathrm{cool}$ is moved from the hot halo to the gaseous disk.  The hot halo gas density is assumed to follow a $\beta$-model profile $\rho(r)= \rho_0 \left(1 + \left(r / r_\beta\right)^2 \right)^{-3 \beta/2}$,
where $r_\beta$ is fixed to 30\% of the virial radius of the halo, $\beta = 2/3$, and $\rho_0$ is determined so that the hot halo has the correct total mass.
Metallicity-dependent cooling curves are computed using \textsc{Cloudy} v23.01\citep{2013RMxAA..49..137F, 2023RMxAA..59..327C} assuming collisional ionisation equilibrium.

\subsection{Star formation}

Star formation in disks is modelled using the prescription of \citet{2006ApJ...650..933B}, which determines the star formation rate surface density from the molecular gas surface density. The total gas surface density depends only upon the gas mass of the disk and the disk radius, while the molecular gas fraction is determined by the midplane hydrostatic pressure of the galactic disk. Higher pressures lead to higher molecular fractions, with the midplane pressure dependent on both the gas and stellar surface densities and their vertical velocity dispersions.

Star formation also takes place within the spheroid, which is modeled as a Hernquist \citep{1990ApJ...356..359H} profile, at a rate $\dot{M}_\star = 0.1 \, M_{\rm gas}/t_{\rm dyn} (V \, / \, 200 \, {\rm km/s})^{-2}$, where $M_{\rm gas}$ is the spheroid gas mass, $t_{\rm dyn}$ is the dynamical time of the spheroid at its Hernquist scale radius and $V$ its circular velocity at the same radius.

\subsection{Metal enrichment}
We track metal enrichment using the instantaneous recycling approximation, wherein stars with masses greater than $1 \msun$ (and therefore with lifetimes shorter than $\sim 10 \, \mathrm{Gyr}$) are assumed to evolve and return (metal enriched) gas to the interstellar medium instantaneously. We adopt a \citet{2001ApJ...554.1274C} initial mass function, with a recycled fraction of $R=0.46$ and a metal yield of $p=0.035$. Metals are assumed to be well-mixed within their associated gas at all times.

\subsection{Stellar feedback}

In addition to returning enriched material to the surrounding gas, star formation also injects energy into the surrounding gas through feedback processes, primarily associated with supernova explosions. Stellar feedback is modelled in a simple manner, with the \emph{mass loading factor} \citep[the mass of gas expelled per unit mass of stars formed, see e.g][]{2020MNRAS.494.3971M} scaling as a power-law in the characteristic velocity. Specifically, the gas outflow rate is
\begin{equation}
\dot{M}_{\mathrm{outflow}} = \left( \frac{V}{V_{\mathrm{outflow}}} \right)^{-\alpha_{\mathrm{outflow}}} \dot{M}_{\star},
\label{eq:powerLawFeedback}
\end{equation}
separately for both the disk and spheroid. Here, $\dot{M}_{\star}$ is the star formation rate in the respective component (disk or spheroid), $\dot{M}_{\mathrm{outflow}}$ is the rate of gas flow from the respective component to the \emph{outflowed} component (see Section \ref{sect:gasRecycling}), $V$ is the characteristic velocity of the component, and $V_\mathrm{outflow}$ is a characteristic velocity scale associated with the stellar feedback. In this work, $\alpha_{\mathrm{outflow}}$ is always positive, such that $V_\mathrm{outflow}$ is the velocity scale above which supernovae are no longer efficient at ejecting material.

For feedback due to star formation in the spheroid we adopt $V_\mathrm{outflow} = 100 \kms$ and $\alpha_\mathrm{outflow} = 2$, while we vary the parameters controlling supernova feedback in the disk in our MCMCs, labelling them $V_\mathrm{disk}$ and $\alpha_\mathrm{disk}$ respectively.

\subsection{AGN feedback}


Supermassive black holes are seeded at the centre of every newly formed dark matter halo, with an initial mass of $100 \msun$ and zero spin. These supermassive black holes then accrete gas from both the hot halo and the spheroid at rates proportional to the  Bondi-Hoyle-Lyttleton rate \citep{1944MNRAS.104..273B}, boosted by factors of $6$ and $5$ for accretion from the hot halo and spheroid respectively, and limited to the Eddington rate, $\dot{M}_{\rm Edd}$.  In addition, supermassive black holes grow through merging, with the black holes assumed to merge instantaneously (i.e. black holes merge when their host galaxies do).

For the purpose of calculating the liberated feedback energy, the gas accretion onto the black holes can take two distinct forms. For both low ($\dot{M}_{\rm acc} < 0.01 \, \dot{M}_{\rm Edd}$) and high ($\dot{M}_{\rm acc} > 0.3 \, \dot{M}_{\rm Edd}$) accretion rates, accretion proceeds through an advection dominated accretion flow (ADAF) \citep{1994ApJ...428L..13N}, while for intermediate accretion rates, there is a radiatively-efficient, geometrically thin, \citet{1973A&A....24..337S} accretion disk. 

For thin disks and high-accretion ADAF phases, the radiative efficiency of accretion is computed as $\epsilon_{\rm rad} = 1 - E_{\rm ISCO}$, where $E_{\rm ISCO}$ is the dimensionless binding energy at the innermost stable circular orbit, determined by the black hole spin. At low accretion rates, corresponding to low-efficiency ADAF states, the radiative efficiency decreases linearly with $\dot{M}_{\rm acc}$, such that it smoothly connects to the thin disk value at $\dot{M}_{\rm acc} = 0.01 \dot{M}_{\rm Edd}$.

The spin of each black hole evolves as it accretes mass and merges with other black holes, which in turn influences both radiative and jet feedback efficiencies. Jet efficiencies for thin disks are spin-dependent, and come from interpolating between non-rotating and rapidly spinning black holes, following the work of \citet{2001ApJ...548L...9M}. For ADAF accretion states, the jet efficiency is determined using the model of \citet{2009MNRAS.397.1302B}.

AGN feedback in \galacticus takes two forms. In the so-called quasar mode, high accretion rates drive winds from the spheroid component of the galaxy. This occurs when the interstellar medium density exceeds a critical threshold \citep{2009ApJ...699...89C}, at which point energy is added to the spheroid gas following the model of \citet{2010ApJ...722..642O}. In the radio mode, the power from black hole jets couples to the hot halo gas, heating it and suppressing cooling. The total efficiency with which the AGN jet power heats the halo gas is governed by a dimensionless coupling parameter $\epsilon_\mathrm{AGN}$. The default value for this parameter is $1$, indicating that all of the calculated jet power is deposited into the surrounding gas. In this work we vary this parameter in our MCMCs, allowing the AGN feedback to be more or less effective than in the default model.

This AGN feedback model is similar in approach to that of the \textsc{Shark} semi-analytic model \citep{2024MNRAS.531.3551L} in that is uses the state of the AGN accretion disk to determine the type, and power of AGN feedback, thereby allowing AGN feedback to occur as a result of \emph{any} source of AGN fueling (e.g. accretion from the hot halo, or from dense gas driven toward the centre of the galaxy as the result of a major merger). This differs from many earlier implementations of AGN feedback in semi-analytic models \citep[e.g.][]{2006MNRAS.370..645B} in which the mode of feedback (``radio'' or ``quasar'' mode) was determined by the source of the AGN fuel. The approach used in \galacticus is more directly motivated by AGN accretion physics (although several key aspects of the model, such as the efficiency with which AGN jets couple to the surrounding hot gas, are described purely by empirical parameters which will eventually require calibration to observations or replacement by more physical models), and is more similar to approaches often implemented into hydrodynamical simulations \citep[e.g.][]{2018MNRAS.479.4056W}.

\subsection{Gas recycling}
\label{sect:gasRecycling}

Gas removed from galaxies by supernova or AGN-driven winds is retained in a reservoir of outflowed gas. This gas then gradually transitions to the hot halo, using the \citet{2013MNRAS.431.3373H} model for this rate of gas recycling: 
\begin{equation}
\dot{M}_\mathrm{return} = \gamma \, (1 + z)^{-\delta_1} \left( \frac{V_\mathrm{vir}}{200\,\mathrm{km\,s}^{-1}} \right)^{\delta_2} \frac{M_\mathrm{outflowed}}{\tau_\mathrm{dyn}},
\label{eq:henriques_reincorporation}
\end{equation}
where $\gamma$, $\delta_1$ and $\delta_2$ are dimensionless parameters, $V_\mathrm{vir} = \sqrt{G M_\mathrm{vir} / r_\mathrm{vir}}$ is the virial velocity of the halo, $M_\mathrm{outflowed}$ is the mass of gas that has been ejected from the disk and/or spheroid by feedback processes, $\tau_\mathrm{dyn}$ is the halo dynamical time and $\dot{M}_\mathrm{return}$ is the rate of mass transfer from $M_\mathrm{outflowed}$ to $M_\mathrm{hot}$. We note that this gas recycling model is not the \galacticus default, but that we chose to use it here because \citet[][]{2013MNRAS.431.3373H} had found that it enabled them to simultaneously match the stellar mass function across a range of redshifts, given the physical models and parameter values chosen in their SAM.

\subsection{Disk and spheroid sizes}
\label{sect:sizes}

The sizes of the disks and spheroids of \galacticus galaxies are found by solving for equilibrium, meaning that their size is such that the angular momentum of the relevant component provides rotational support. This method of determining galaxy sizes was introduced in \citet{2000MNRAS.319..168C}.

In this work, one of the \galacticus parameters associated with spheroid sizes is allowed to vary during our MCMCs, and so here we describe further details of the spheroid size calculation, noting that disk sizes are solved for in a similar manner. Our spheroids are assumed to have \citet{1990ApJ...356..359H} density profiles,
\begin{equation}
\rho(r) = \frac{M}{2\pi} \, \frac{r_\mathrm{s}}{r (r + r_\mathrm{s})^3},
\label{eq:hernquistProfile}
\end{equation}
where $r_\mathrm{s}$ is the Hernquist scale radius, and $M$ is the total spheroid mass.

The starting point for calculating the spheroid size is the spheroid mass, $M$, and angular momentum, $L$,\footnote{Following \citet[Appendix~C]{2000MNRAS.319..168C}, by ``angular momentum''
we mean the \emph{pseudo–angular momentum} of the spheroid: although a
pressure-supported spheroid has little net angular momentum, $L$ is the angular
momentum the system would have \emph{if} it were rotationally supported. For spheroids formed in mergers we determine the size by energy conservation as in
\citet{2000MNRAS.319..168C}; during disk instabilities we assume that the pseudo angular momentum gained by the spheroid is equal to the angular momentum lost by the disk.}
which are both tracked by \galacticus. We define the specific angular momentum of material to be $j$, and the specific angular momentum of the whole spheroid to be $J = L/M$.

The specific angular momentum of material on a circular orbit at a radius of $r_\mathrm{s}$ is $j_\mathrm{s} = v_\mathrm{s} r_\mathrm{s}$, where $v_\mathrm{s}$ is the circular velocity at $r_\mathrm{s}$, which would be $\sqrt{GM_\mathrm{tot}(<r_\mathrm{s})/r_\mathrm{s}}$ if everything was circularly symmetric, where $M_\mathrm{tot}(<r_\mathrm{s})$ is the total mass (disk, spheroid and dark matter halo) within $r_\mathrm{s}$. Writing both the spheroid angular momentum and mass as integrals over radius, the specific angular momentum of the entire spheroid can be written as 
\begin{equation}
J = \left. \int_0^\infty 4 \pi r^2 \rho(r) v(r) r \, \mathrm{d}r \middle/ \int_0^\infty 4 \pi r^2 \rho(r) \, \mathrm{d}r  \right. .
\label{eq:J_integrals}
\end{equation}
With a change of variables to $x = r/r_\mathrm{s}$, and the approximation that the rotation velocity is constant with radius (and equal to $v_\mathrm{s}$) we get that
\begin{equation}
J = r_\mathrm{s} v_\mathrm{s} \left. \int_0^\infty x^3 f(x) \, \mathrm{d}x \middle/ \int_0^\infty x^2 f(x) \, \mathrm{d}x  \right. \equiv r_\mathrm{s} v_\mathrm{s} \, R_3 / R_2,
\label{eq:implicit_rs}
\end{equation}
where $f(x) = 1 \, / \, x(1+x)^3$ is a dimensionless form of the Hernquist density profile, and $R_m$ is the $m^\mathrm{th}$ moment of the dimensionless Hernquist profile. Because $v_\mathrm{s}$ depends upon $r_\mathrm{s}$ equation~\ref{eq:implicit_rs} is not an explicit equation for $r_\mathrm{s}$, but it can be solved iteratively, which is done at the same time as iteratively solving for the disk size and adiabatic contraction of the dark matter halo by the baryon distribution.

The Hernquist profile's 3rd moment, $R_3$, logarithmically diverges, because the integrand goes as $1/x$ at large $x$. This implies that a Hernquist profile would have infinite angular momentum if the rotation velocity is constant. In practice, the circular velocity will not be exactly constant, and is expected to drop at sufficiently large radii, resolving the problem of an infinite total angular momentum. However, rather than finding $L$ for the Hernquist spheroid by using $v(r)$ calculated from the total matter density profile (adding to the complication of the coupled equations that we are seeking to solve iteratively), we continue to assume that  $v(r) = v_\mathrm{s}$, but treat $R_2 / R_3$ as a free parameter of the model. The default value used by \galacticus for $R_2 / R_3$ is 0.5, which is approximately the value it would have if the moment integrals that enter $L$ and $M$ were truncated at $6 \, r_\mathrm{s}$. We allow the value of $R_2 / R_3$ to vary during our MCMC runs.

\section{Model calibration}
\label{sect:calibration}

As discussed in the Introduction, a common dataset to try to fit when calibrating a SAM is the low-redshift stellar mass function. This is popular both because it can be precisely measured, and because the total stellar mass is one of the most fundamental properties of a galaxy. However, the sophisticated physical models in \galacticus make it computationally expensive to directly calibrate to the stellar mass function, because of the large number of halos that need to be simulated for a model prediction of the stellar mass function.

\subsection{Calibration strategy}
\label{sect:calibrationStrategy}

The stellar mass function, $\Phi(M_\star)$, describes the comoving number density of galaxies per unit stellar mass.\footnote{Note that in the relevant figures we express $\Phi$ as a number density per logarithmic interval of stellar mass.} It can be expressed as an integral over the halo mass function, $\Phi_\mathrm{h}(M_\mathrm{h})$, convolved with the probability distribution, $P(M_\star|M_\mathrm{h})$, for a halo of mass $M_\mathrm{h}$ to host a galaxy of stellar mass $M_\star$. This leads to the expression
\begin{equation}
\label{eq:SMF_from_SHMR}
\Phi(M_\star) = \int \Phi_\mathrm{h}(M_\mathrm{h}) \, P(M_\star|M_\mathrm{h}) \, \mathrm{d}M_\mathrm{h}.
\end{equation}

If the SHMR relation is deterministic (that is, all halos of a particular halo mass will host a galaxy with a particular stellar mass), then the integral simplifies to a change of variables,
\begin{equation}
\Phi(M_\star) = \Phi_\mathrm{h}(M_\mathrm{h}) \left| \frac{\mathrm{d}M_\mathrm{h}}{\mathrm{d}M_\star} \right|.
\label{eq:SMF_from_HMF}
\end{equation}
In practice, we expect scatter in the SHMR, such that galaxies with a range of stellar masses can all live in halos with a particular mass. By trying to jointly fit the observed stellar mass function and galaxy clustering, it appears that this scatter is well described as approximately lognormal, with a scatter of around 0.2 dex in the stellar mass at fixed halo mass, with no obvious mass trend \citep{2013ApJ...771...30R}. We note that because the halo mass function is rather steep (with roughly an order of magnitude more halos per decade of halo mass as one goes down an order of magnitude in halo mass), the scatter in the SHMR has an appreciable effect on the stellar mass function. At a particular stellar mass, there are many more low mass halos that can ``scatter up'' to that stellar mass, than there are halos for which this stellar mass would be typical. As such, the scatter in $P(M_\star|M_\mathrm{h})$ typically increases the stellar mass function compared with that of a model with the same median SHMR but no scatter.

Given that the cosmological parameters are tightly constrained \citep{2020A&A...641A...1P}, and that -- at least in this work -- we are trying to calibrate our SAM at fixed cosmology, we can assume that we know $\Phi_\mathrm{h}(M_\mathrm{h})$ for the real Universe with high precision. The form of Equation~\ref{eq:SMF_from_SHMR} then suggests an approach to reproducing the observed stellar mass function: reproduce the observationally inferred SHMR, encoded in $P(M_\star|M_\mathrm{h})$. This change in calibration target is useful because the halo mass is something over which we have complete control, whereas the stellar mass is a property that arises from running \galacticus. As such, we can target particular halo masses and measure the \galacticus prediction for $P(M_\star|M_\mathrm{h})$ at those masses, whereas we cannot easily target a particular $M_\star$ to measure the \galacticus prediction for $\Phi(M_\star)$ there.

\subsection{Calibrating to the low-$z$ SHMR}
\label{sect:calibratingLowzSHMR}

As a demonstration of our new calibration strategy, we first fit only to the low-redshift $(z \sim 0.3)$ SHMR. With the eventual goal of calibrating a model that would be useful for populating large volume cosmological simulations with \galacticus galaxies, we use a mass resolution for the merger trees of $2 \times 10^{10} \msun$. This is similar to the mass resolution of the merger trees from the UNIT simulations \citep{2019MNRAS.487...48C} that have previously been used with \galacticus for generating mock catalogues for the Roman telescope's High Latitude Spectroscopic Survey \citep{2019MNRAS.490.3667Z, 2021MNRAS.501.3490Z}.

For our target SHMR, we use the results from \citet[][hereafter \ltw]{2012ApJ...744..159L}, which constrain the SHMR in three redshift bins between $z=0.22$ and $z=1$ using a joint analysis of galaxy–galaxy weak lensing, galaxy clustering, and the stellar mass function. Initially, we fit only to the lowest-redshift bin. We target halo masses of $\log_{10}(M_\mathrm{h}/\msun) \approx 11.5$, 12.0, and 12.5 at $z=0.29$. Going to lower halo masses would require merger trees below our $2\times10^{10}\,\msun$ mass resolution, while higher masses become increasingly expensive to evolve with \galacticus. These three masses bracket the \emph{knee} of the SHMR, where the slope changes from stellar-feedback dominated to AGN-feedback dominated \citep[e.g.][]{1991ApJ...379...52W, 2006MNRAS.365...11C}, allowing us to constrain both regimes simultaneously.

In \ltw, the SHMR represents the stellar–halo mass relation of \emph{central} galaxies as defined within their halo–occupation framework. Centrals and satellites are not identified on a galaxy-by-galaxy basis. Instead, a parametric form for the central SHMR (with scatter) is inferred statistically by jointly fitting the lensing, clustering, and number-density measurements. Thus, the halo masses in \ltw\ are not measured individually, nor does the analysis require determining which observed galaxies are centrals or satellites.

We note that our choice of \ltw as a target SHMR is primarily driven by convenience, because a comparison with this relation (taking into account the appropriate weighting of different redshifts, differences in halo virial mass definition, and differences in assumed cosmology between \galacticus and the analysis of the observations) was already available within \galacticus. More recent work has found results that are reasonably consistent with those from \ltw, but has extended the relationship out to higher redshifts \citep{2022A&A...664A..61S}. This suggests that our results would not change substantially if we used a more recent determination of the SHMR, although a comparison with more recent work would be useful if we want to better constrain the model to match the high-redshift Universe.

The \ltw relation that we calibrate to is only for \emph{central} galaxies, meaning that it does not include satellites of more massive galaxies. This makes generating the model prediction considerably easier, because we have complete control of the halo masses of central objects, whereas the (sub)halo masses of satellites depend on processes such as dynamical friction and tidal stripping, that take place within \galacticus. The downside to only considering centrals is that the properties of observed satellites are not being directly calibrated to, meaning that physics that primarily affects the evolution of satellites (tidal stripping, ram pressure stripping, etc.) will not be constrained (beyond any subtle effects they have on central galaxies). We note that none of the parameters that we leave free in our MCMCs are primarily concerned with the evolution of satellites, and that this situation is not all that different from calibrating to only the stellar mass function above $\sim 10^9 \msun$, which is dominated by central objects.\footnote{Roughly 20\% of galaxies with $M_\star > 10^{9} \msun$ are satellites, dropping to around 5\% for galaxies with $M_\star > 10^{11} \msun$ \citep[see Fig.~8 from][]{2022A&A...664A..61S}.}

\subsubsection{SHMR likelihood}

We use three halo-mass bins indexed by $b\in\{1,2,3\}$, with each bin having a single target halo mass.  
Each bin contains $N_b$ halos (typically $N_b=9$). We define the log stellar mass for the galaxy in halo $i$ from bin $b$, 
\begin{equation}
\mathcal{M}^{\mathrm{sim}}_{bi}\;\equiv\;\log_{10}\!\left(\frac{M_{\star,bi}}{\msun}\right).
\end{equation}
The estimated mean log stellar mass and scatter in a particular bin is then
\begin{align}
\big\langle \mathcal{M} \big\rangle_b^{\mathrm{sim}}
&= \frac{1}{N_b}\sum_{i=1}^{N_b}\mathcal{M}^{\mathrm{sim}}_{bi},
\label{eq:meanM_sim}\\
\left(\sigma_b^{\mathrm{sim}}\right)^2
&\equiv \frac{1}{N_b-1}\sum_{i=1}^{N_b}\!\left(\mathcal{M}^{\mathrm{sim}}_{bi}-\big\langle \mathcal{M} \big\rangle_b^{\mathrm{sim}}\right)^2,
\label{eq:sigma_sim}
\end{align}
with the standard errors (i.e. uncertainties) on these quantities being
\begin{align}
\delta\!\big(\langle \mathcal{M} \rangle_b^{\mathrm{sim}}\big)
  &= \frac{\sigma_b^{\mathrm{sim}}}{\sqrt{N_b}}, \label{eq:sem}\\
\delta\!\big(\sigma_b^{\mathrm{sim}}\big)
  &= \frac{\sigma_b^{\mathrm{sim}}}{\sqrt{2\,(N_b-1)}}.\label{eq:stderr_sigma}
\end{align}

We define $\big\langle \mathcal{M} \big\rangle_b^{\mathrm{obs}}$ and $\sigma_b^{\mathrm{obs}}$ to be the SHMR mean and intrinsic scatter from \ltw, with uncertainties $u_b$ and $v_b$ respectively. \ltw measure a scatter in inferred stellar mass at fixed halo mass of 0.21 dex. However, this scatter is a combination of intrinsic scatter and measurement uncertainties. As explained in \ltw, the photometric redshifts that they use lead to considerably larger stellar mass uncertainties than when the SHMR is measured from spectroscopic data. As such we take $\sigma_b^{\mathrm{obs}} = 0.16\,\mathrm{dex}$ and $v_b=0.04\,\mathrm{dex}$, for all $b$, as advocated for in \ltw \citep[based on the finding of][]{2009MNRAS.392..801M}.

Using the definitions above, we can write down $\chi^2$-like terms for how well the \galacticus galaxies match the observationally-inferred mean SHMR and its scatter
\begin{align}
\chi^2_{\mathrm{mean}} &= \sum_{b=1}^{3}
\frac{\left(\big\langle \mathcal{M} \big\rangle_b^{\mathrm{sim}} - \big\langle \mathcal{M} \big\rangle_b^{\mathrm{obs}}\right)^2}
     {\left(\sigma_b^{\mathrm{sim}}\right)^2/N_b + u_b^2}, \label{eq:chi2_mean} \\[0.5em]
\chi^2_{\mathrm{scat}} &= \sum_{b=1}^{3}
\frac{\left(\sigma_b^{\mathrm{sim}} - \sigma_b^{\mathrm{obs}}\right)^2}
     {\left(\sigma_b^{\mathrm{sim}}\right)^2/[2\,(N_b-1)] + v_b^2}.
\end{align}

For our MCMC the product of the likelihood and prior is proportional to the posterior probability that the MCMC samples from. Our prior is listed in Table~\ref{tab:galacticus_priors}, which also serves as a description of the parameters varied during our MCMCs. For our likelihood, we use (up to some multiplicative constant)\footnote{
While MCMC does not require a normalised posterior density (it depends only on relative posterior probabilities), strictly, one should include the likelihood normalisation when the covariance varies, because the Gaussian normalisation term $\tfrac{1}{2}\ln\det(2\pi\Sigma)$ then changes. For example, $\chi^2_{\mathrm{mean}}$ is small not only when $\langle \mathcal{M} \rangle_b^{\mathrm{sim}}$ matches $\langle \mathcal{M} \rangle_b^{\mathrm{obs}}$, but also for parameter choices that produce a larger simulated scatter in $\mathcal{M}^{\mathrm{sim}}_{bi}$ (since $\sigma_b^{\mathrm{sim}}$ appears in the denominator of Eq.~\ref{eq:chi2_mean}). In practice this effect is small here because we also fit the observed scatter in the SHMR, so credible posterior samples have similar $\sigma_b^{\mathrm{sim}}$, rendering the normalisation approximately constant.
} 
\begin{equation}
\label{eq:likelihood_from_chi2}
\mathcal{L} = \exp\left( -\frac{1}{2} \left(\chi^2_\mathrm{mean} / T_\mathrm{mean} + \chi^2_\mathrm{scat}  / T_\mathrm{scat} \right) \right), 
\end{equation}
where $T_\mathrm{mean} = 3$ and $T_\mathrm{scat} = 3$ are dimensionless ``temperatures'' with which we heat up the different components of the likelihood.

This heating was done because our primary goal is to find \galacticus models that are a good fit to observational data, rather than being concerned with the constraints on the \galacticus parameters themselves. Heating up the likelihood reduces the probability of our MCMC sampler getting ``stuck'' in local posterior maxima. Also, \galacticus is only an approximate model of galaxy formation, and so we do not expect it to be able to match all observational data ``down to the noise level''. As such, the amount of weight that we choose to give to matching different observables is fairly arbitrary and can be chosen by us, rather than being dictated by  how precisely different properties of observed galaxies are measured. For example, we may wish to try and match both the number density of galaxies of a particular stellar mass (i.e. the stellar mass function) and the mean size of those galaxies to similar fractional precision, even though observationally one of these things may be more precisely determined than the other. Appropriately heating the different contributions to the likelihood allows us to choose what weight we give to different constraints.

The choice of $T_\mathrm{mean} = T_\mathrm{scat} = 3$ was fairly arbitrary, but was motivated by practical considerations related to the numerical behaviour of the MCMC and the level of accuracy we consider meaningful for the present calibration. In particular, the intrinsic scatter in log stellar mass at fixed halo mass is typically $\sim 0.15$--$0.2$\,dex. With nine haloes per mass bin, this implies an expected uncertainty on the mean log stellar mass for a given \galacticus realisation of $\sim 0.05$--$0.07$\,dex. Heating the likelihood by a factor of three, which is equivalent to inflating the effective uncertainties by $\sqrt{3}$, increases this scale to $\sim 0.1$\,dex. We regard this as an appropriate target precision given the approximate nature of the model and the fact that our primary goal is to identify regions of parameter space that provide broadly acceptable fits to the data, rather than to derive formally precise constraints on individual model parameters.


\begin{table*}
\centering
\begin{tabular}{llll}
\toprule
\textbf{Parameter} & \textbf{\galacticus name} & \textbf{Prior} & \textbf{Description} \\
\midrule
$ V_\mathrm{disk} \, / \, \mathrm{km \, s^{-1}} $ & \tt{velocityCharacteristic} & $\ln \mathcal{N}(150, 0.5; 25, 300)$ & Characteristic velocity for disk outflows (eq.~\ref{eq:powerLawFeedback}) \\
$ \alpha_\mathrm{disk} $ & \tt{exponent} & $\mathcal{U}(0.0, 5.0)$ & Power-law slope for disk outflows (eq.~\ref{eq:powerLawFeedback}) \\
$ \gamma $ & \tt{gamma} & $\ln\mathcal{N}(5.0, 0.5; 1, 25)$ & Reincorporation timescale normalization (eq.~\ref{eq:henriques_reincorporation}) \\
$ \delta_1 $ & \tt{delta1} & $\mathcal{U}(-2.0, 4.0)$ & Redshift dependence of reincorporation timescale (eq.~\ref{eq:henriques_reincorporation}) \\
$ \delta_2 $ & \tt{delta2} & $\mathcal{U}(-1.0, 5.0)$ & Halo mass dependence of reincorporation timescale (eq.~\ref{eq:henriques_reincorporation}) \\
$ \epsilon_\mathrm{AGN} $ & \tt{efficiencyRadioMode} & $\ln\mathcal{N}(0.3, 0.5; 0.03, 1)$ & Efficiency of AGN heating in radio mode \\
$ R_2 / R_3 $ & \tt{ratioAngularMomentumScaleRadius} & $\ln \mathcal{N}(0.2, 0.5; 0.1, 0.5)$ & Controls spheroid sizes (Section~\ref{sect:sizes}) \\
\bottomrule
\end{tabular}
\caption{Priors used for the \galacticus parameters in our MCMC analysis. The \textbf{\galacticus name} is the parameter name as it appears in \galacticus configuration files. For parameters with uniform priors, $\mathcal{U}(a,b)$ denotes a uniform distribution between $a$ and $b$. For parameters with lognormal priors, $\ln \mathcal{N}(x_0, \sigma; a, b)$ denotes a lognormal distribution with median $x_0$ and dispersion $\sigma$ in the natural logarithm of the parameter (so $\sigma = 0.5$ corresponds to a width of 0.22 dex), sharply truncated below $a$ and above $b$.}
\label{tab:galacticus_priors}
\end{table*}

\subsubsection{Differential evolution MCMC}
\label{sect:differentialEvolution}

We use the differential evolution MCMC algorithm \citep[DE-MC,][]{2006S&C....16..239T} to perform our MCMCs. In DE-MC, a population of parallel chains is evolved, with the proposed step for each chain being a scaled difference between randomly chosen members of the population. This adapts both the proposal scale and orientation to the shape of the posterior distribution, leading to efficient sampling without requiring prior knowledge of the shape of the posterior. We note that DE-MC is the same method as used by \citet{2011MNRAS.416.1949L} to calibrate their SAM. 

\begin{figure}
        \centering
        \includegraphics[width=0.49\textwidth]{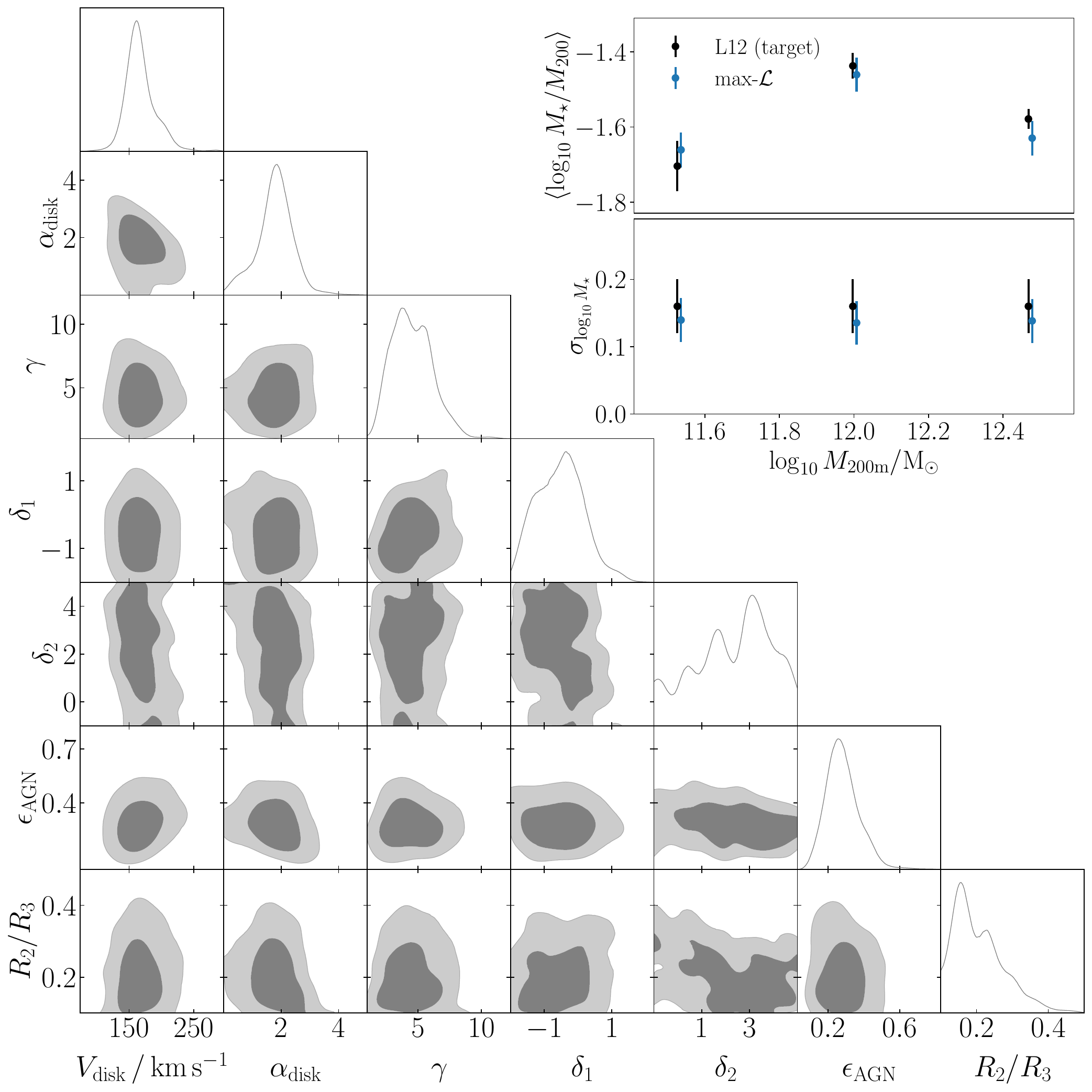} 
	\caption{A corner plot showing the posterior distribution for the \galacticus parameters that we varied, when fitting only to the SHMR (and its scatter) at $z \approx 0.3$. The panels in the top-right show the target data from \ltw, as well as the corresponding model data vector for the maximum-likelihood model from our MCMC.}
	\label{fig:SHMRcorner}
\end{figure}

For the first demonstration of our method (fitting only to the low-$z$ SHMR and its scatter) we ran DE-MC with 16 parallel chains, with each chain being of length 466, of which we removed the first 47 steps as burn-in. A corner plot of the resulting posterior distribution is shown in Fig.~\ref{fig:SHMRcorner}, with the target data from \ltw and the best-fitting model plotted in the top-right panel. 

The integrated autocorrelation times \citep{Sokal1997} of our chains vary between about 13 and 20 steps for the different parameters. This means that across all of our 16 chains, we have an effective sample size of only a few hundred samples. If we were primarily interested in the posterior distribution of \galacticus parameters, then we would want to run considerably longer chains (to reduce sampling noise) and remove a longer burn-in phase (to remove any bias from the initial conditions of our chains). However, for our purpose of just finding a model of \galacticus that is ``good enough'' for the purpose of generating future mock galaxy catalogues, the fact that our posteriors are not fully converged is not a concern. 

\subsubsection{Sample variance in the likelihood}

Simulating only 27 halos for a likelihood evaluation means that our calculated likelihood for a given model is subject to sample variance. If we consider a single mass bin, then we have 9 halos all with the same mass. The scatter (standard deviation) in $\log_{10} M_\star$ at fixed $M_\mathrm{halo}$ is approximately 0.15 dex, so with 9 halos we expect the scatter on the mean log stellar mass, $\left< \log_{10} M_\star \right>$, to be $\sim 0.15 / \sqrt{9} = 0.05 \, \mathrm{dex}$. If our goal was to find a model that matched the SHMR to better precision than 0.05 dex then we would need to use more halos in each likelihood evaluation to ensure that sample variance in the model prediction does not significantly bias our inferred parameters. However, as demonstrated later (see Section~\ref{sect:results_fullCalibration} and Fig.~\ref{fig:fullCalibrationMCMC}) when trying to fit to multiple properties (as opposed to just the SHMR at a single redshift) we typically ``miss'' the SHMR by $\gtrsim 0.1 \, \mathrm{dex}$ anyway, and so a 0.05 dex model uncertainty on $\left< \log_{10} M_\star \right>$ for a particular set of \galacticus parameters has negligible effect in practice.

When doing inference with only a small number of simulated merger trees, there is a choice to make: should the trees be kept fixed for each likelihood evaluation, or should we generate a new set of random trees for each likelihood evaluation? Fixed trees would mean that the particular details of the trees used might systematically bias the results, while random trees would lead to a ``noisy'' likelihood, which would macroscopically vary over arbitrarily small parameter changes. 

For an example of the bias with fixed trees, consider that there is both theoretical and observational evidence that the expected stellar mass at fixed halo mass depends on the formation time of a halo \citep{2017ApJ...836..161L, 2023ApJ...959....5L}. Taking as a simple caricature of this ``earlier forming halos have higher stellar masses'', then if our sample of 9 merger trees were early forming (compared with the full population of merger trees with this final halo mass) the mean stellar mass in our likelihood evaluation for a particular \galacticus model would be higher than the actual (zero sample variance) \galacticus prediction for this model. This would then bias us towards favouring galaxy formation physics parameters that reduce the stellar masses in these halos.

New random trees for each likelihood evaluation would mean that very similar (or even identical) \galacticus parameters would produce different sets of galaxies, and therefore a different likelihood. This increases the probability of an MCMC walker getting ``stuck'' because a particular likelihood evaluation had a fortuitous set of merger trees, that led it to fit the data better than it would have done with a more typical set of trees, which in turn decreases the probability of the MCMC accepting newly proposed states.

While many MCMC algorithms can be successfully applied in cases where one has a noisy estimate of the likelihood \citep{Alquier2014NoisyMC}, we nevertheless sought to avoid problems associated with this, and so opted to use a fixed set of trees throughout our MCMCs. If one wished to reduce the bias associated with any particular realisation of merger trees, the most straightforward approach would be to increase the number of trees per likelihood evaluation. One could, however, imagine more efficient alternatives: for example, motivated by the discussion above, at fixed halo mass one could select merger trees so as to evenly sample the distribution of halo formation times, rather than relying on purely random draws. Such a deliberately constructed set of trees could reduce sample variance for a fixed number of simulated halos. We leave exploration of such approaches to future work, stressing again that in practice we do not find a model that fits all our calibration data with precision comparable to our model sample variance.


\subsubsection{Posterior distribution of \galacticus parameters}
\label{sect:posterior_lowzSHMRcalibration}

Figure~\ref{fig:SHMRcorner} shows the posterior distributions of the \galacticus parameters when calibrating only to the low-redshift SHMR. There is a slight anti-correlation between the disk outflow normalisation parameter $V_\mathrm{disk}$ and its power-law slope $\alpha_\mathrm{disk}$. This behaviour is expected, as increasing either parameter raises the mass-loading factor of stellar feedback in low-mass galaxies (see equation~\ref{eq:powerLawFeedback}), and similar SHMR predictions can therefore be obtained through compensating changes in these two parameters. By contrast, the AGN feedback efficiency parameter $\epsilon_\mathrm{AGN}$ shows little correlation with the other parameters in this calibration.

The parameters governing gas reincorporation, $\gamma$, $\delta_1$, and $\delta_2$, which control the normalisation and the redshift and halo-mass dependence of the reincorporation timescale, are only weakly constrained by the low-redshift SHMR alone, with $\delta_2$ in particular being poorly determined. This indicates that, within the present calibration, variations in the reincorporation timescale have only a modest effect on the stellar masses of central galaxies at fixed halo mass. To explore this further, we experimented with varying these parameters by hand and found that their impact on galaxy properties becomes substantially more pronounced if the rate at which gas cools from the hot halo to the disk is artificially increased. This suggests that, for the models considered here, the dominant bottleneck for recycled gas returning to the star-forming disk is the cooling time of the hot halo rather than the reincorporation timescale itself, rendering the recycling parameters subdominant in this particular calibration. 
We note that this behaviour differs from that found in the semi-analytic models of \citet{2013MNRAS.431.3373H,2015MNRAS.451.2663H}, where delaying the reincorporation of ejected gas was shown to play an important role in regulating the stellar mass growth of low-mass galaxies. This difference may reflect variations in the treatment of gas cooling and halo thermodynamics between models, which can alter whether reincorporation or cooling constitutes the dominant bottleneck for gas returning to the star-forming disk. We defer a more detailed investigation of this interplay between cooling and recycling physics to future work, where additional observables and/or alternative cooling prescriptions can be investigated.

\subsubsection{Results of model calibrated to the low-$z$ SHMR}
\label{sect:results_lowzSHMRcalibration}
    
\begin{figure*}
        \centering
        \includegraphics[width=0.49\textwidth]{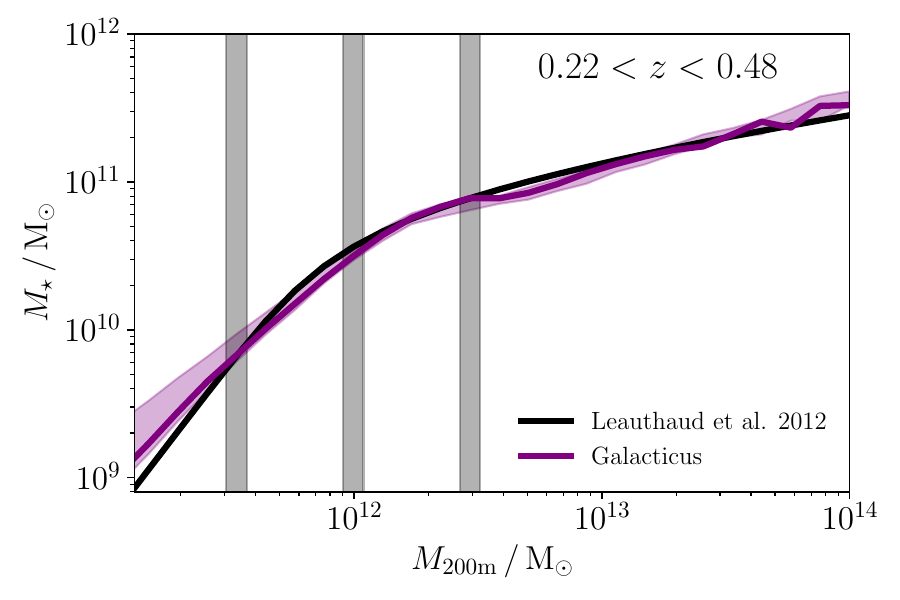} 
        \includegraphics[width=0.49\textwidth]{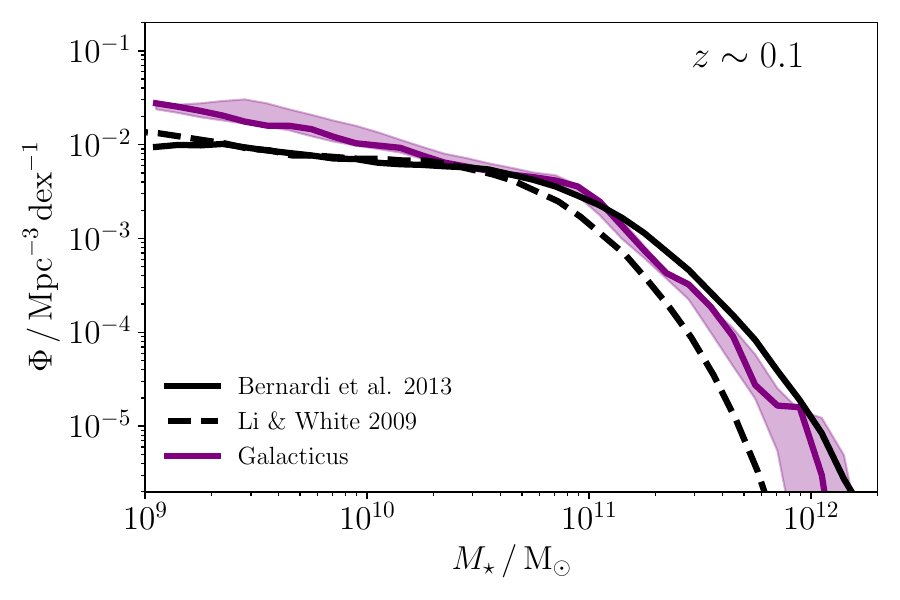} 
	\caption{The low-redshift SHMR (left) and stellar mass function (right) from a \galacticus model calibrated only to the low-$z$ SHMR in three mass bins (see Section~\ref{sect:results_lowzSHMRcalibration}). The mass bins used for calibration are marked by the grey vertical lines in the left panel. The solid lines show the \galacticus results for our MAP parameters, while the shaded regions mark the 16th--84th percentiles, evaluated by running \galacticus with parameters randomly sampled from the MCMC chains.}
	\label{fig:SHMRcalibration}
\end{figure*}

In Fig.~\ref{fig:SHMRcalibration} we show the results of taking the \galacticus model parameters inferred in Fig.~\ref{fig:SHMRcorner} and running \galacticus with these parameters on a larger number of merger trees spanning a range of halo masses. We simulate 30,000 halos with masses between $10^{11} \msun$ and $10^{14} \msun$, with the distribution of masses following the \citet{2002MNRAS.329...61S} halo mass function with slightly updated numerical parameters taken from \citet{2019MNRAS.485.5010B}. The resulting SHMR and stellar mass function for the maximum a posteriori (MAP) \galacticus parameters are shown by the solid lines in Fig.~\ref{fig:SHMRcalibration}, with the shaded region covering the 16th--84th percentiles from running \galacticus with random draws from the posterior distribution (shown in Fig.~\ref{fig:SHMRcorner}).

In contrast to the calibration stage, which uses galaxies evaluated at a single output redshift ($z=0.29$), the comparisons shown in Fig.~\ref{fig:SHMRcalibration} account for the finite redshift width of the observational data. When comparing \galacticus predictions to the \ltw\ SHMR, we therefore do not rely on a single model output time. Instead, the model SHMR is constructed by combining galaxies from multiple \galacticus outputs whose associated redshift intervals overlap the observational bin, with each output weighted by the comoving volume of its overlap with the observational redshift range.

As expected, the model is a good match to the SHMR data at the halo masses that featured in the likelihood used for calibration. In addition, the model does a good job at reproducing the SHMR even at halo masses considerably higher than those calibrated to, with a posterior distribution for $\left< \log_{10} M_\star / \msun \right>$ at a halo mass of $10^{13.5} \msun$ seemingly as tight as at the halo masses used for calibration. It is not obvious that this should be the case, but this success perhaps reflects the fact that the galaxies in these more massive halos have built up a large fraction of their stellar mass through mergers \citep[e.g.][]{2016MNRAS.458.2371R} such that their total stellar mass depends more upon star formation in lower mass halos (where the model is being calibrated) than on star formation in massive halos themselves.

We now turn to the implied stellar mass function. Unlike the SHMR, which is inferred over a fixed redshift interval, the SDSS stellar mass function is derived from a magnitude-limited sample \citep{2009ApJS..182..543A}, such that galaxies of different stellar masses are observable over different redshift ranges. To account for this, the \galacticus predictions are constructed by combining galaxies from multiple output times, with each output weighted by the comoving volume over which galaxies of a given stellar mass would be detectable in the survey.

At the massive end, published SDSS stellar mass functions show significant systematic differences. In particular, \citet{2013MNRAS.436..697B} find a higher abundance of massive galaxies than \citet{2009MNRAS.398.2177L}. As discussed by \citet{2013MNRAS.436..697B}, this difference arises from a combination of factors, including a more complete treatment of extended low–surface-brightness stellar envelopes and the mitigation of background over-subtraction in large galaxies that had biased stellar masses low in earlier analyses.

Our success in matching the SHMR at high halo masses leads to a stellar mass function that is in reasonable agreement with observations at the massive end. The calibrated \galacticus model falls between the determinations of \citet{2013MNRAS.436..697B} and \citet{2009MNRAS.398.2177L}. The offset relative to \citet{2013MNRAS.436..697B} is expected, as \galacticus does not explicitly model a separate stellar halo or intrahalo light component, and therefore reports stellar masses for the inner galaxy component only. Conversely, one could argue that because \galacticus does not model tidal stripping of stars from satellite galaxies prior to their final merger, stellar mass that would contribute to an extended stellar halo in other models (and in the real Universe) remains bound to satellites until coalescence (ultimately contributing to the inner galaxy). However, satellite merger timescales can be long, such that a substantial fraction of satellites do not contribute their stellar mass to the central galaxy by $z \sim 0.1$. As a result, we expect the stellar masses reported by \galacticus to most closely correspond to only the inner regions of observed galaxies.

Where the model differs more significantly from observations is at the low mass end, with the number density of \galacticus galaxies with $10^9 < M_\star / \msun < 10^{10}$ over-predicting that seen in the data by a factor of roughly two. This reflects the fact that \galacticus's $M_\star(M_{200})$ is too high at low halo masses, as well as being ``too flat''. This causes a wider range of halo masses to produce galaxies within some stellar mass range, increasing their number density (as can be seen mathematically in equation~\ref{eq:SMF_from_HMF}). This deficiency could potentially be addressed by including a further halo mass bin at lower masses in the calibration, though we do not pursue this here because the fixed resolution of our merger trees means that these halos' merger histories would be poorly resolved.


\subsection{Including additional observational constraints}
\label{sect:includingAdditionalConstraints}

While the low-redshift stellar mass function has often been treated as a benchmark for galaxy formation models, it is certainly not the only property of the observed galaxy population that one might wish their model to reproduce. In the context of SAMs, many other properties have been calibrated to, with examples including gas fractions and the stellar mass--metallicity relation \citep{2015MNRAS.453.4337S}, the black hole mass to bulge mass relation \citep{2015ApJ...801..139R}, and the fraction of quiescent galaxies and the neutral hydrogen mass function \citep{2025arXiv250415283A}.

The exact properties that one chooses to use for calibrating a model will depend upon the use cases for the calibrated model. In the future we hope to use the techniques presented here to calibrate \galacticus to produce mock galaxy catalogues for upcoming large scale structure surveys such as will be done by Roman and Rubin. For now however, our model calibration is as a proof of concept, so we choose a couple of additional datasets to calibrate to that highlight some of the complications that arise when not simulating a large sample of halos for each likelihood evaluation. Specifically: we demonstrate how to extend our approach to calibrate to data at multiple redshifts, by including a higher-redshift SHMR in the calibration data,  and we show how non-SHMR data can be calibrated to, using the galaxy stellar mass--size relation as an example.

\begin{figure*}
        \centering
        \includegraphics[width=0.65\textwidth]{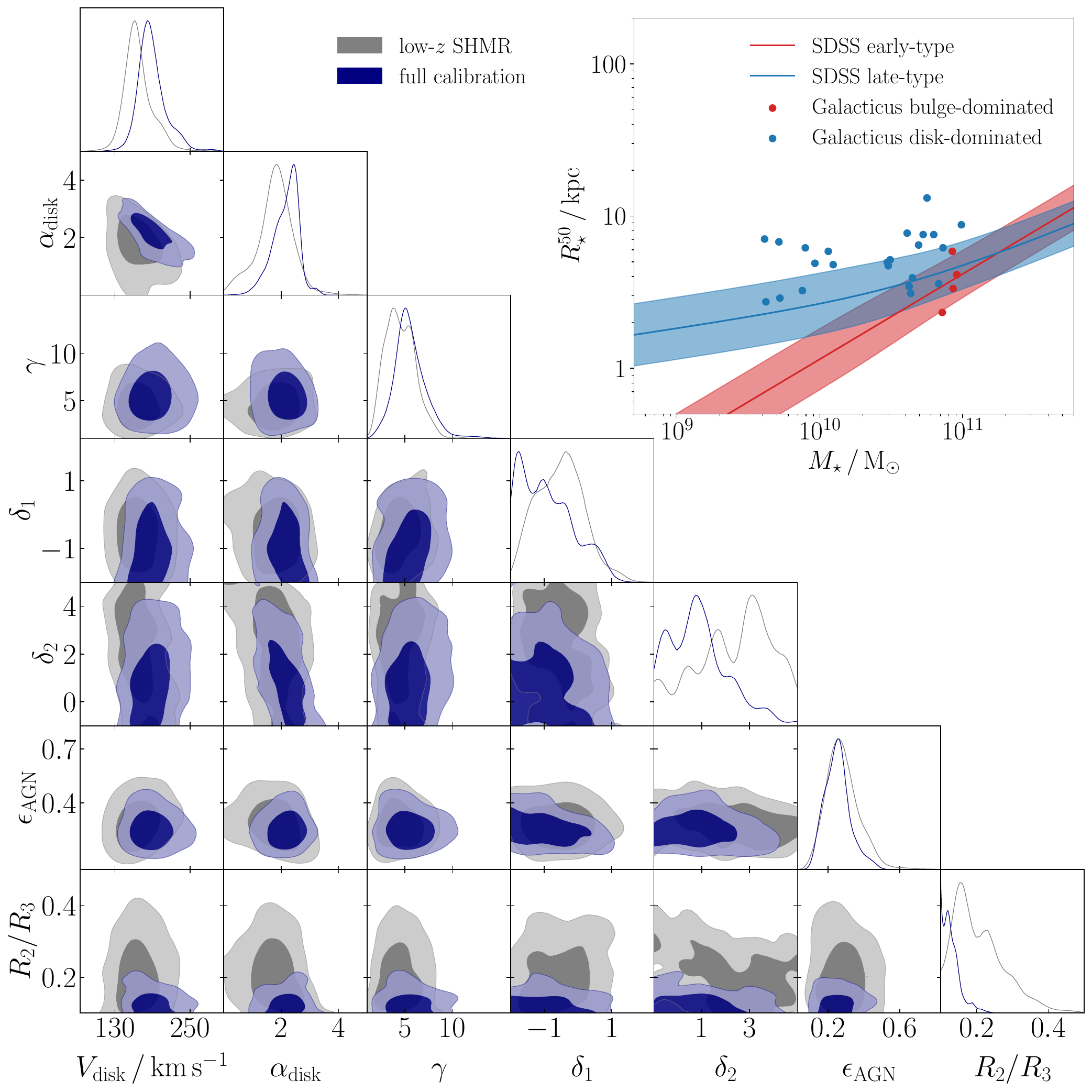} 
        \includegraphics[width=0.34\textwidth]{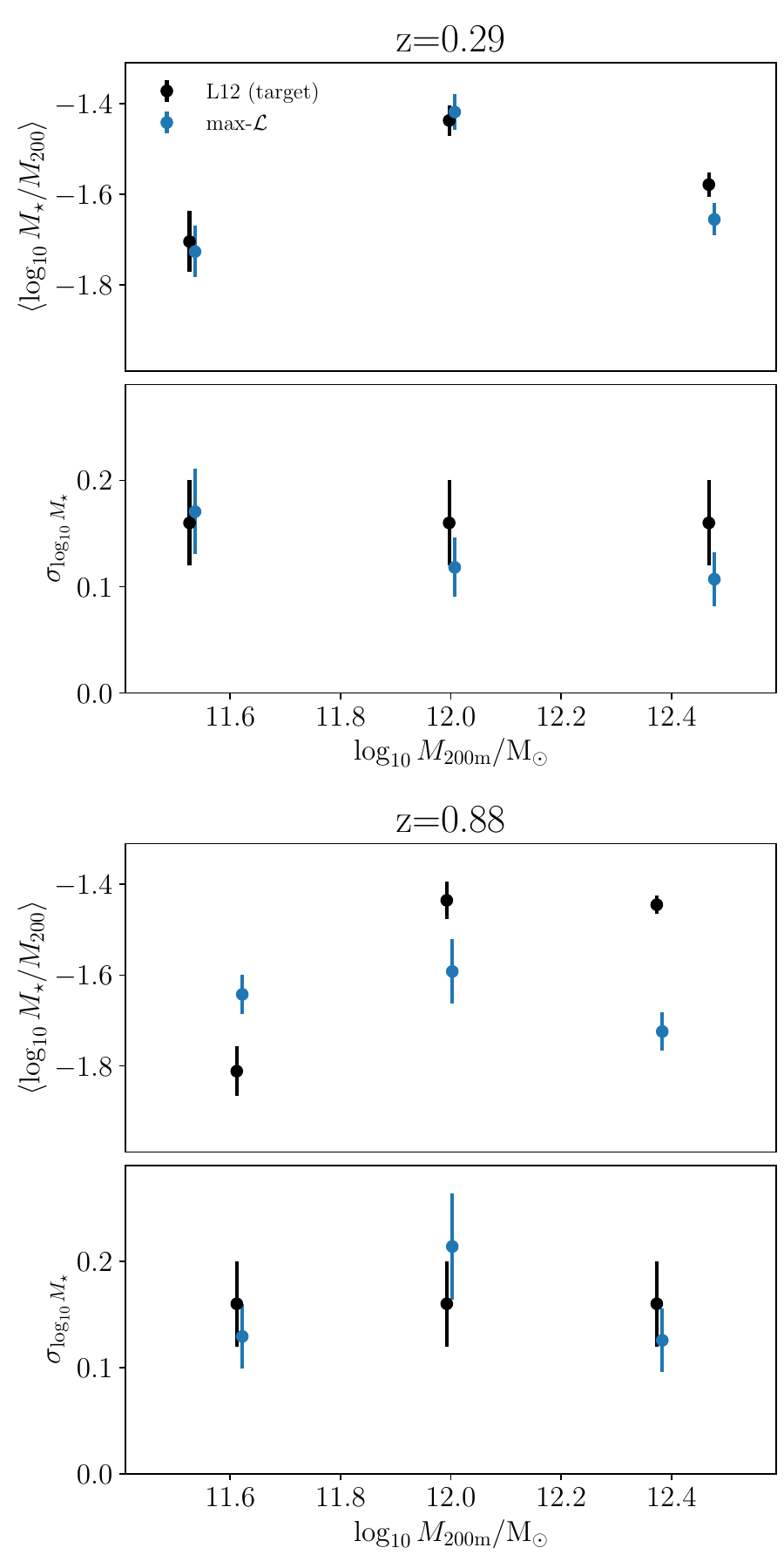} 
	\caption{A corner plot showing the posterior distribution for the \galacticus parameters that we varied, when fitting to the SHMR (and its scatter) at $z \approx 0.3$ and $z \approx 0.9$, as well as to the low-redshift stellar mass--size relation. This posterior is labelled ``full calibration'', with the posterior distribution from Fig.~\ref{fig:SHMRcorner} also shown for comparison here, and labelled ``low-$z$ SHMR''. The rightmost panels are similar to the inset panel in Fig.~\ref{fig:SHMRcorner}, for both the low and high redshift SHMR. The $M_\star$--$R_\star^{50}$ panel shows the relationship between galaxy stellar mass and galaxy half-light radius for the maximum-likelihood model from our MCMC chains. The median target relations for both early-type and late-type galaxies and the associated lognormal scatter are shown by the lines with shaded regions \citep[from][]{2003MNRAS.343..978S}. The dots mark the locations of 27 \galacticus galaxies, with the colour denoting whether they have a $B/T>0.5$ (which we associated with early-type galaxies), or $B/T \leq 0.5$ (which we associated with late-type galaxies). These 27 galaxies are the same 9 halos per mass bin over three mass bins as used for the low-$z$ SHMR.}
	\label{fig:fullCalibrationMCMC}
\end{figure*}



\subsubsection{The SHMR at higher redshift}

The goal of many upcoming LSS surveys is to extend measurements of galaxy clustering to the higher-redshift Universe. As such, we would like our calibrated model to not only match the properties of local galaxies, but of galaxies in the more distant Universe also. \ltw calculated a SHMR for three different redshift bins, the $0.22 < z < 0.48$ bin used in Section~\ref{sect:calibratingLowzSHMR}, a $0.48 < z < 0.74$ bin, and a $0.74 < z < 1$ bin. Here, we additionally fit to the $0.74 < z < 1$ SHMR from \ltw, which we will sometimes refer to as ``high-redshift''.

One drawback of our approach of targeting specific halo masses at low-redshift is that these simulated halos do not then target specific halo masses at higher redshift. In contrast, if one simulates a representative sample of halos at $z=0$, then the ensemble of their progenitors forms a statistically representative sample of halos at earlier epochs (i.e. reproducing the halo mass function above the resolution limit). This means that to apply the same technique as we used for the low-$z$ SHMR at multiple redshifts, we need to simulate a bespoke set of merger trees for each target redshift. Much like calibrating to a few specific halo masses with the hope that the physically-motivated nature of SAMs would produce results that also match observations at other masses (as seen in Fig.~\ref{fig:SHMRcalibration}), it seems reasonable that matching galaxy properties at a small number of different redshifts would also lead to galaxy properties being reasonable at intermediate redshifts not directly calibrated to. As such, while we note that our method would become inefficient if matching galaxy properties at many different redshifts, we suspect that in practice this will not be particularly important. This assumption effectively relies on the model behaving sensibly when interpolated between the calibrated redshifts. That is not to say that we necessarily expect it to perform well when extrapolated far beyond the calibrated redshift range: as seen for the SHMR in Fig.~\ref{fig:SHMRcalibration}, where the best-fitting parameters do not extend accurately to lower halo masses.

Similar to the low-$z$ SHMR we target three halo masses, $\log_{10} M_\mathrm{h} / \msun = 11.6$, 12.0 and 12.4, now at a redshift of 0.88. These are again chosen to bracket the knee of the SHMR. The slight change in the halo masses simulated compared with the low-$z$ case is primarily driven by the fact that the \ltw relation at high-$z$ only goes down to a halo mass of $10^{11.6} \msun$.

\subsubsection{The stellar mass--size relation}
\label{sect:mass_size_likelihood}

As previously mentioned, there are many properties of observed galaxies that one would ideally like a SAM to reproduce. In practice, one must choose a limited set of observables to use as explicit calibration targets, and our method is naturally best suited to observables that correlate directly with halo mass, since halo mass is the quantity we control when generating merger trees. An obvious example of such an observable -- which we focus on here -- is the size of galaxies, which has long been predicted within SAMs and compared to observations.

One potential avenue for extending our method to calibrate \galacticus using galaxy sizes would be to fit to an inferred halo mass--size relation. While the connection between halo mass and galaxy size is considerably less well studied than the SHMR, there has been work in this direction \citep{2017ApJ...838....6H, 2020MNRAS.492.1671Z}. However, these studies typically associate observed galaxies with halos via an assumed SHMR, such that the resulting halo mass--size relation is effectively a reparameterisation of the stellar mass--size relation. With this in mind, we choose to calibrate \galacticus directly to the observed stellar mass--size relation.

If we had a large number of simulated galaxies for each likelihood evaluation, then the size distribution as a function of stellar mass could be compared between observations and \galacticus. In our case, however, we only have 27 (central) galaxies, with hand chosen halo masses. We therefore opt to use a likelihood which is a product of a separate term for each \galacticus galaxy, based upon how consistent that galaxy's size is with the size distribution (given its stellar mass) in the observed Universe.

For target data we use the \citet{2003MNRAS.343..978S} stellar mass--size relation, based on 140,000 galaxies from the Sloan Digital Sky Survey (SDSS). \citet{2003MNRAS.343..978S} split galaxies into early-type (i.e. spheroid-dominated elliptical galaxies) and late-type (disk-like galaxies) based on a cut in S\'ersic index. They then measured the median and lognormal scatter in the half-light radii of galaxies, $R_\star^{50}$, in bins of stellar mass, finding that these were well-described by simple functional forms. The median relations and scatter are shown in Fig.~\ref{fig:fullCalibrationMCMC}.

For each central galaxy that forms, we first decide whether to treat it as an early-type or late-type galaxy. We approximate this by making a cut on the bulge-to-total stellar mass ratio, $B/T = M_\star^\mathrm{bulge} / (M_\star^\mathrm{bulge} + M_\star^\mathrm{disk})$, with $B/T < 0.5$ being classified as disk-dominated/late-type, and  $B/T \geq 0.5$ being classified as bulge-dominated/early-type. In practice, the distribution of $B/T$ values in \galacticus is strongly disk-dominated at low stellar masses and bimodal at high stellar masses, such that any threshold in the range $0.4 \lesssim B/T \lesssim 0.8$ would change the assigned morphological type of only a small fraction of galaxies and would not materially affect the results presented here. The morphological classification determines with which of the two \citet{2003MNRAS.343..978S} relations we will compare this galaxy's size. The galaxy's stellar mass, $M_{\star,bi}$, and morphological type, $t_{bi}$, determine $p_\mathrm{obs}^{(t_{bi})}(R_\star^{50} | M_{\star,bi})$ from \citet{2003MNRAS.343..978S}. The product of this probability density (evaluated at the half-light radius of the galaxy in question, $R_{\star,bi}^{50}$) gives us the mass--size component of the likelihood, 
\begin{equation}
\label{eq:sizesLikelihood}
\mathcal{L}_\mathrm{size}
= \prod_{b=1}^{3}\;\prod_{i=1}^{N_b}
p_{\mathrm{obs}}^{(t_{bi})}\!\big(R_{\star,bi}^{50}\mid M_{\star,bi}\big).
\end{equation}

We evaluate the observational density $p_{\mathrm{obs}}^{(t_{bi})}\!\big(R_{\star,bi}^{50}\mid M_{\star,bi}\big)$ at the simulated points, i.e.\ we ask how plausible the model outputs would be if drawn from the observational distribution. This is distinct from the probability of the observed galaxy sizes, given a particular \galacticus model, so we regard it as a quasi-likelihood. Subtleties associated with this distinction are discussed in Appendix~\ref{app:size_quasilike}.


While we have focussed here on galaxy sizes, we are imagining that this method of calibrating a SAM to galaxy properties beyond just stellar masses could be used for other properties also. For example, one could imagine constraining to the stellar mass--black hole mass relation \citep[such as from][]{2013ApJ...764..184M} in a similar manner. Calibrating to multiple datasets has been an important part of improving the fidelity of SAMs \citep[e.g.][]{2015ApJ...801..139R}, and so we would ideally want a rigorously justified likelihood that we could use to enable this, when using a small number of simulated galaxies in each likelihood evaluation. We are unaware of a way to do this, though for now it appears that the method we have presented here may suffice.

\subsubsection{MCMC with combined constraints}

We vary the same \galacticus\ parameters with the same priors as in Sec.~\ref{sect:calibratingLowzSHMR} (Table~\ref{tab:galacticus_priors}). Our tempered joint likelihood is then
\begin{multline}
\label{eq:joint_like}
\ln \mathcal{L}_\mathrm{joint} =
-\tfrac{1}{2}\!\Bigg[
\frac{\chi^2_{\mathrm{mean}}(z_\mathrm{low})}{T_{\mathrm{mean}}^{\mathrm{low}}}
+\frac{\chi^2_{\mathrm{scat}}(z_\mathrm{low})}{T_{\mathrm{scat}}^{\mathrm{low}}}
\\
+\frac{\chi^2_{\mathrm{mean}}(z_\mathrm{high})}{T_{\mathrm{mean}}^{\mathrm{high}}}
+\frac{\chi^2_{\mathrm{scat}}(z_\mathrm{high})}{T_{\mathrm{scat}}^{\mathrm{high}}}
\Bigg]
+\frac{1}{T_{\mathrm{size}}}\,\ln \mathcal{L}_{\mathrm{size}} .
\end{multline}
In our runs we set $T_{\mathrm{mean}}^{\mathrm{low}}=T_{\mathrm{scat}}^{\mathrm{low}}
=T_{\mathrm{mean}}^{\mathrm{high}}=T_{\mathrm{scat}}^{\mathrm{high}}=3$ and $T_{\mathrm{size}} = 9$.

The resulting posterior distribution is shown in Fig.~\ref{fig:fullCalibrationMCMC}. Overall, the posterior is qualitatively similar to that obtained when fitting only to the low-$z$ SHMR, although some shifts are evident when constraints at multiple redshifts and on galaxy sizes are included. In particular, there is a modest preference for stronger stellar feedback (higher $V_\mathrm{disk}$) with a steeper halo-mass dependence (higher $\alpha_\mathrm{disk}$).

The most pronounced tightening of the posterior occurs for $R_2/R_3$, as expected given the inclusion of galaxy size data in the likelihood. In addition, the gas reincorporation parameters ($\gamma$, $\delta_1$, and $\delta_2$) exhibit shifts relative to the low-$z$-only calibration. That $\delta_1$, which controls the redshift dependence of gas reincorporation, is affected by the inclusion of multi-epoch constraints is physically intuitive, although the parameter remains only weakly constrained. The posteriors for $\gamma$ and $\delta_2$, which together regulate the normalisation and halo-mass dependence of the reincorporation timescale, shift modestly but remain weakly constrained. These changes may reflect degeneracies between gas cycling timescales, stellar feedback, and galaxy sizes, since the timing of gas return to the disk can affect both stellar mass growth and the angular momentum of subsequently formed stars.\footnote{In \galacticus, gas cooling onto the disk is assumed to inherit the specific angular momentum of the host halo at the time of cooling, so changes in gas cycling timescales can indirectly affect galaxy sizes by shifting the epoch at which gas settles into the disk.}

Fig.~\ref{fig:fullCalibrationMCMC} also shows how the best-fitting model compares with the calibration data. The model generally performs well, although disk sizes are, on average, slightly too large, and the $z\sim0.9$ SHMR is too flat, with stellar masses that are too high at low halo mass and too low at high halo mass.


\subsubsection{Calibrated quantities over a wider range of halo masses}
\label{sect:results_fullCalibration}

\begin{figure*}
    \centering
    \includegraphics[width=0.49\textwidth]{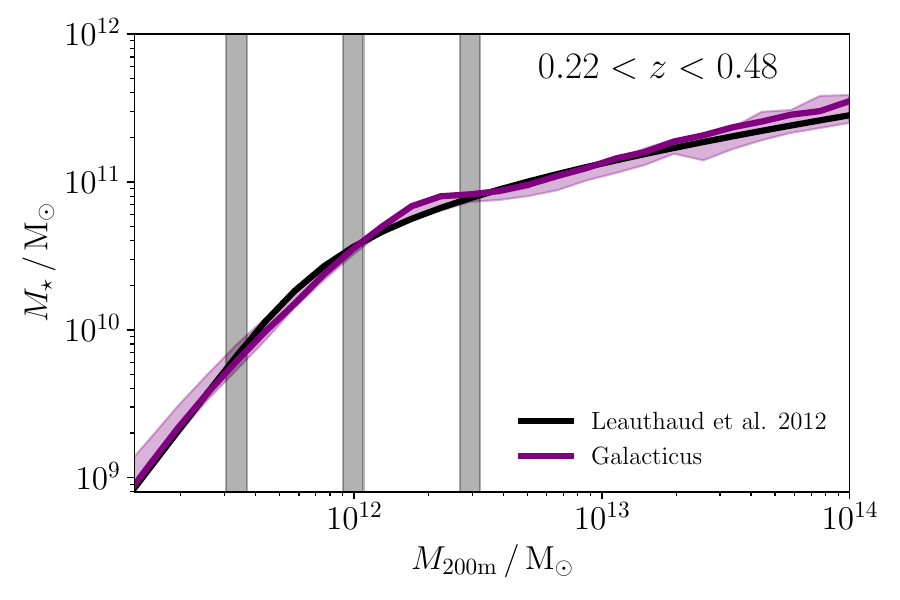}
    \includegraphics[width=0.49\textwidth]{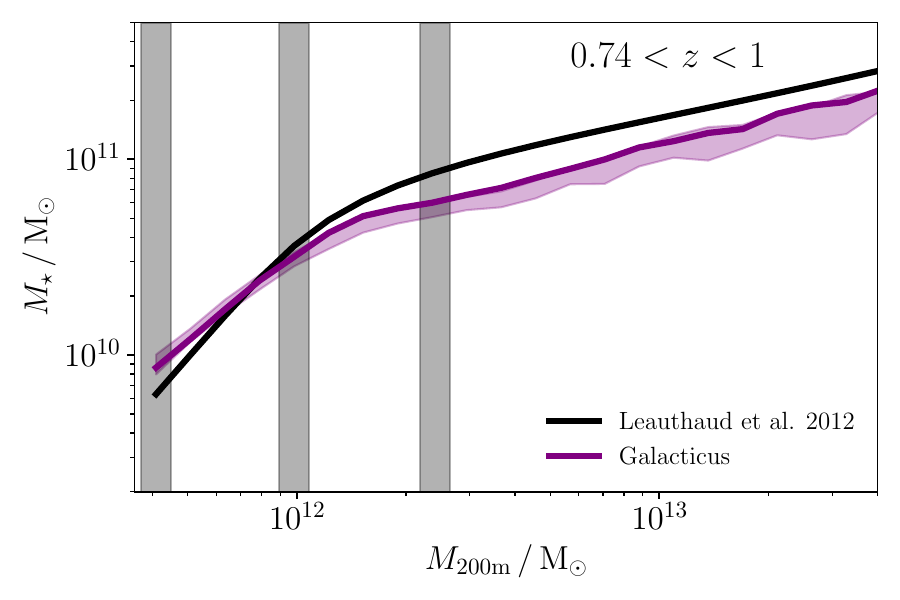}
    \caption{Low-$z$ (left) and high-$z$ (right) SHMRs for the \galacticus model calibrated to the low-$z$ and high-$z$ SHMR and the low-$z$ stellar mass--size relation (see Section~\ref{sect:includingAdditionalConstraints}). The mass bins used for the SHMR calibration are marked by the grey vertical lines in each panel. The solid lines show the \galacticus results for our MAP parameters, while the shaded regions mark the 16th--84th percentiles, obtained by running \galacticus with parameters randomly drawn from the MCMC chains. Despite being calibrated to the high-$z$ data, the model does not reproduce a sufficiently steep SHMR at high-$z$ to match the \ltw measurements.}
    \label{fig:SHMR_lowz_and_highz}
\end{figure*}

Before turning to broader diagnostics in Section~\ref{sect:results}, we assess how the calibrated model performs against the very observables used for calibration, and how it extrapolates beyond the halo-mass range directly constrained by the likelihood. To do this, we take our MAP parameters and use these to simulate 40,000 central halos, with masses following the halo mass function from \citet{2019MNRAS.485.5010B}, over the mass range $10^{11}$ to $10^{15} \msun$.

\medskip
\noindent
\textbf{SHMR at low-$z$ and high-$z$.} At $z \sim 0.3$ (left panel of Fig.~\ref{fig:SHMR_lowz_and_highz}) the model reproduces the SHMR well across the calibrated bins and extrapolates reasonably towards lower halo masses. This apparent improvement relative to the model calibrated only to the low-$z$ SHMR (compare with Fig.~\ref{fig:SHMRcalibration}) should be interpreted cautiously, as the mass accretion histories of our lowest-mass halos are not well resolved. Nonetheless, it hints that the physics required to obtain realistic galaxy sizes at fixed halo mass may also help regularize the stellar mass growth of lower-mass systems.

At $z\sim 0.9$ (right panel), however, the calibrated model is too shallow compared to the \ltw\ SHMR, and the shortfall persists even in the very halo-mass bins used during calibration (see also Fig.~\ref{fig:fullCalibrationMCMC}). This indicates that the current combination of \galacticus\ sub-grid prescriptions and the subset of parameters we vary does not provide enough freedom to simultaneously match the low-$z$ SHMR, the high-$z$ SHMR, and the local stellar mass--size relation.

\begin{figure}
    \centering
    \includegraphics[width=0.49\textwidth]{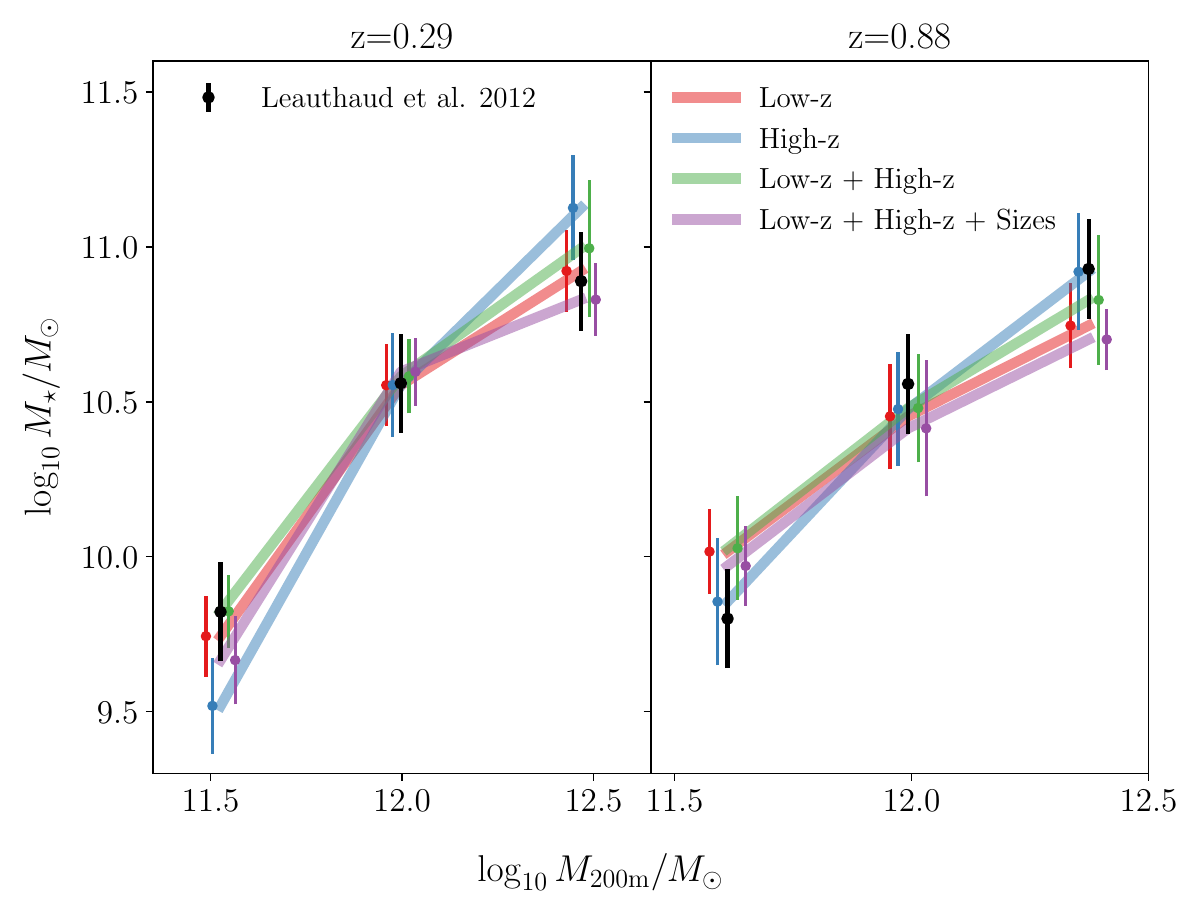}
    \caption{Tensions encountered when fitting multiple datasets. Black points with error bars show the \ltw\ means and scatters in $\log M_\star$ at $z=0.29$ (left) and $z=0.88$ (right) used in the likelihood. Coloured symbols show the best-fitting models when calibrating to different dataset combinations (legend). Points are offset slightly in $M_{200\mathrm{m}}$ for clarity; coloured lines connect the unshifted model points to aid comparison.}
    \label{fig:matchingMultipleDatasets}
\end{figure}

The trade-offs are illustrated in Fig.~\ref{fig:matchingMultipleDatasets}. Calibrating to only the low-$z$ SHMR yields an excellent match at $z \sim 0.3$ but a too-flat relation at higher redshift; calibrating to only the high-$z$ SHMR achieves the correct steepness at $z\sim 0.9$ but over-steepens the relation at $z \sim 0.3$. Jointly fitting both redshifts leads to a compromise that is intermediate between the two single-redshift fits. Adding the \citet{2003MNRAS.343..978S} mass--size relation to the likelihood further shifts the MAP model towards lower stellar masses at high halo mass, resulting in a $z\sim 0.9$ SHMR that is considerably too flat.

We note that part of the tension illustrated in Fig.~\ref{fig:matchingMultipleDatasets} may reflect limitations of the observational constraints themselves, rather than shortcomings of the physical prescriptions in Galacticus alone. The SHMR measurements at different redshifts, and the low-redshift stellar mass–size relation, are derived from heterogeneous datasets and methodologies, and are affected by systematic uncertainties (e.g. stellar population modelling, IMF assumptions, dust corrections, and size definitions) that are not fully captured by the quoted statistical errors. As a result, these datasets are not guaranteed to be mutually consistent. In such circumstances, no single parameter set may reproduce all constraints simultaneously within their nominal uncertainties, even if the underlying model is broadly reasonable. While this likely contributes to the compromises seen here, the residual systematic offsets between model predictions and multiple observables nevertheless motivate additional model flexibility, which we discuss further below.

\begin{figure}
    \centering
    \includegraphics[width=0.49\textwidth]{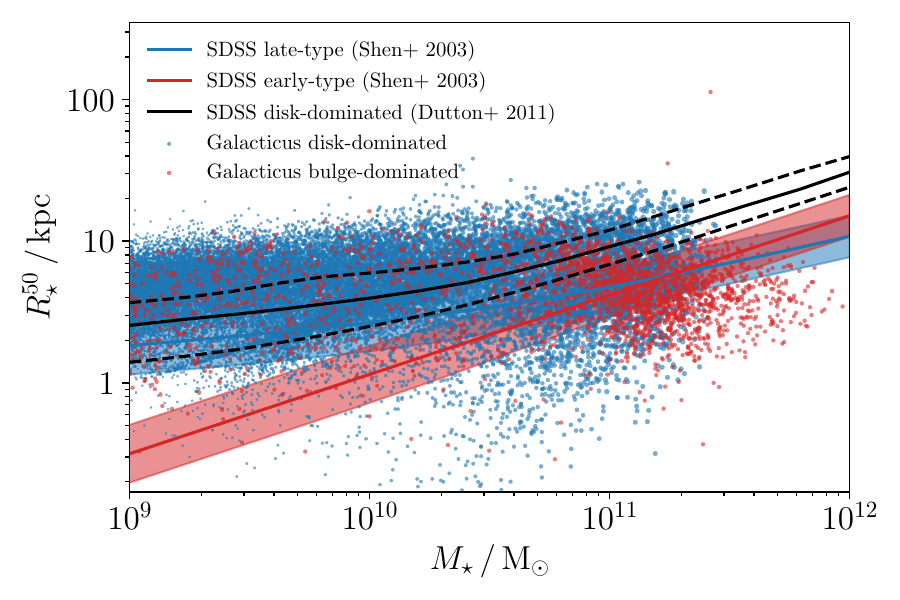}
    \caption{Stellar mass--size relation at $z\simeq0.07$ for the calibrated \galacticus\ model.
SDSS medians and $1\sigma$ scatters from \citet{2003MNRAS.343..978S} are shown for early/late types (red/blue curves with shaded bands).
Model galaxies are plotted as points with the same colour scheme, using a $B/T$-based disk/bulge split.
Model disks are systematically too large and the slope is slightly too shallow; model bulges show little mass dependence in size, at odds with \citet{2003MNRAS.343..978S}.
Part of the disk-size offset reflects size-definition differences: \citet{2003MNRAS.343..978S} report circularized radii, whereas semi-major-axis sizes from \citet{2011MNRAS.410.1660D} reduce the discrepancy.}
    \label{fig:MstarR50}
\end{figure}

\medskip
\noindent
\textbf{Stellar mass--size relation.} We compare the $z \approx 0.07$ $M_\star$--$R^{50}_\star$ relation of the calibrated model to the SDSS measurements from \citet{2003MNRAS.343..978S} in Fig.~\ref{fig:MstarR50}. Despite including these data in the likelihood, our model disks are somewhat too large and the slopes of the $M_\star$--$R^{50}_\star$ relations are too shallow, especially for bulge-dominated systems. 

Part of the apparent disk-size offset reflects differences in size definitions. \citet{2003MNRAS.343..978S} report \emph{circularized} half-light radii, which bias inclined disks to smaller sizes relative to semi-major-axis measurements. When we instead compare to the SDSS re-analysis of \citet{2011MNRAS.410.1660D} -- who use semi-major-axis sizes -- the discrepancy for disks is substantially reduced (see also Section~\ref{sect:size_definitions}). Thus, while our model does not perfectly match the \citet{2003MNRAS.343..978S} relation to which it was calibrated, this could reflect the fact that the \citet{2003MNRAS.343..978S} sizes are unphysically small \citep[for the reasons outlined in][]{2011MNRAS.410.1660D} such that an inability to match them may not represent a failing of our \galacticus model.

For bulge-dominated systems the calibrated model shows an almost mass-independent size, in tension with the observed early-type trend. This likely reflects limited constraining power in our fast likelihood: within the calibration set only a small number of galaxies end up bulge-dominated, and these span only a narrow stellar mass range. We discuss options to improve the size constraints in Section~\ref{sect:improveSizesMatch}.

\section{The results of our calibrated model}
\label{sect:results}

Our calibrated model does not fully reconcile the low-$z$ and high-$z$ SHMRs with the low-$z$ galaxy size distribution. Nevertheless, it represents a substantial improvement over the default \galacticus configuration and over previous \galacticus calibrations developed for large–scale structure mock catalogues \citep{2018MNRAS.474.5206K, 2019MNRAS.490.3667Z}, particularly in reproducing the low-$z$ stellar mass function. We make the details of this calibration publicly available,\footnote{\href{https://doi.org/10.5281/zenodo.16952803}{doi.org/10.5281/zenodo.16952803}} including our model choices, parameter values, and the specific \galacticus version used, with the expectation that future refinements will further improve agreement with observational constraints.

For applications to mock large–scale structure analyses, the two primary cosmological probes are galaxy clustering and gravitational lensing. Gravitational lensing sensitivity depends critically on galaxy brightness and size, which affect shape–measurement accuracy. Clustering measurements require accurate galaxy redshifts; at higher redshifts -- such as those targeted by DESI, Euclid, and Roman -- these are typically obtained via emission–line detection, making realistic emission–line luminosities essential for robust mock catalogues. With these considerations in mind, we present three key comparisons between the galaxy population produced by our model and observations:  
(i) the stellar mass function,  
(ii) the stellar mass–size relation, and  
(iii) the H$\alpha$ emission–line luminosity function.
In addition, we examine the relationship between stellar mass and star formation rate in order to interpret the origin of discrepancies between the simulated and observed emission--line luminosity functions.

\subsection{The stellar mass function through time}

\begin{figure}
        \centering
        \includegraphics[width=0.49\textwidth]{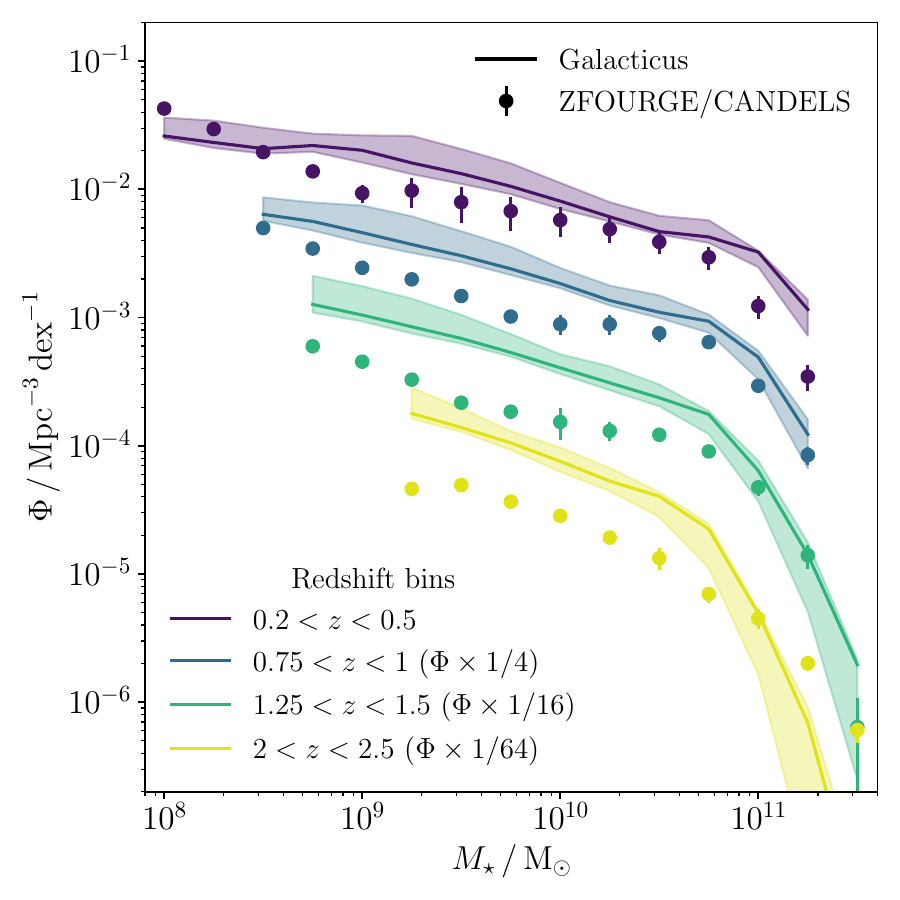} 
	\caption{The stellar mass function of our calibrated \galacticus model (from Section~\ref{sect:includingAdditionalConstraints}) compared with the stellar mass function inferred from ZFOURGE/CANDELS data in \citet{2014ApJ...783...85T}. For clarity, we only show every second redshift bin from \citet{2014ApJ...783...85T}, and shift the mass functions for higher redshift bins down in normalisation, by amounts indicated in the legend. The solid lines show the MAP model, while the shaded regions shows the 2.5th to 97.5th percentile distributions from runs done with parameters drawn from the posterior shown in Fig.~\ref{fig:fullCalibrationMCMC}.}
	\label{fig:ZFOURGEcomparison}
\end{figure}

In Fig.~\ref{fig:ZFOURGEcomparison} we compare the stellar mass function (SMF) predicted by our fully calibrated model (Section~\ref{sect:includingAdditionalConstraints}) to the observational determinations from ZFOURGE/CANDELS presented by \citet{2014ApJ...783...85T}. For clarity, we show only every second redshift bin from the observations, and offset the higher-redshift bins in the vertical direction as indicated in the legend. Importantly, the SMF was \emph{not} a direct calibration target in our likelihood function. Instead, it emerges from constraining the model parameters by the SHMR at low and high redshift, together with the low-redshift stellar mass--size relation. The comparison in Fig.~\ref{fig:ZFOURGEcomparison} therefore provides a demonstration of how well our calibration strategy generalises to a key observable across cosmic time.  

At low redshift ($0.2 < z < 0.5$), the agreement between model and data is generally very good across the observed stellar mass range. This success largely reflects the direct constraint imposed by the $z \sim 0.3$ SHMR (Fig.~\ref{fig:SHMR_lowz_and_highz}), which fixes the normalisation and slope of the SMF near the knee. We note that in comparison to \citet{2014ApJ...783...85T}, we slightly over-predict the number of galaxies, especially at the high-mass end. This is in contrast to the earlier comparison with \citet{2013MNRAS.436..697B}, highlighting the significant systematic uncertainties and differences in stellar mass definitions between observational studies. As discussed earlier, the stellar mass estimates from \citet{2013MNRAS.436..697B} include extended diffuse emission, that is not tracked by \galacticus, nor included in most other observational measurements of the SMF. The comparison here therefore likely represents a genuine (though small) excess of massive galaxies in our model. We note that the \ltw SHMR used for calibration assigns relatively high stellar masses at fixed halo mass compared to other empirical inferences \citep[e.g.][]{2013MNRAS.428.3121M, 2019MNRAS.488.3143B}, which likely contributes to this behaviour.

At intermediate redshifts ($0.75 < z < 1.5$), the model still reproduces the overall shape of the SMF reasonably well, but begins to systematically over-produce low-mass galaxies. This trend mirrors that seen in the SHMR comparison at $z \sim 0.9$ (Fig.~\ref{fig:SHMR_lowz_and_highz}), where the model fails to produce a sufficiently steep relation.  This discrepancy becomes more pronounced in the highest redshift bin, where the model predicts a factor of $\sim3$ excess of low-mass galaxies and a deficit of high-mass systems relative to the \citet{2014ApJ...783...85T} measurements.

Overall, this comparison shows that our SHMR-based calibration method yields a model that reproduces the observed SMF well at $z \lesssim 1$, despite the SMF not being an explicit calibration target. The increasing mismatch at earlier times highlights the limitations of the present model in capturing the redshift evolution of the SHMR, and points to the need for enhanced flexibility (e.g. in feedback, gas recycling, or cooling models). If such flexibility is introduced, it may also be necessary to incorporate additional high-redshift constraints to ensure accurate predictions at $z \gtrsim 1.5$.

\subsection{Emission line luminosities}

We model nebular emission lines in \galacticus using the methodology described in \citet{2018MNRAS.474..177M}, which is based on a method first presented in \citet{2003A&A...409...99P}. In brief, we pre-compute a library of \textsc{Cloudy} \citep{2013RMxAA..49..137F} \ion{H}{2} region models spanning a range of hydrogen densities, metallicities, and ionising spectra. For each galaxy, we generate an integrated stellar spectrum from its star formation and chemical enrichment history, compute the ionising photon output, and estimate characteristic ISM densities and the number of \ion{H}{2} regions in its disk and bulge. By interpolating the ionising flux and ISM properties over the \textsc{Cloudy} grid, we obtain the emission line luminosity per \ion{H}{2} region, which is then scaled by the number of regions to give the total luminosity.

This framework allows \galacticus to predict any emission line included in the \textsc{Cloudy} library in a manner consistent with the galaxy’s star formation and chemical enrichment history. Here, we focus on the $\mathrm{H}\alpha$ line, which is a physically meaningful tracer because its luminosity is closely tied to the instantaneous star formation rate \citep[e.g.][]{1998ARA&A..36..189K}, and because it is a primary feature used to measure galaxy redshifts in ongoing and future space-based spectroscopic surveys: Euclid \citep{2011arXiv1110.3193L} and the Nancy Grace Roman Space Telescope \citep{2015arXiv150303757S}.

\begin{figure}
        \centering
        \includegraphics[width=0.49\textwidth]{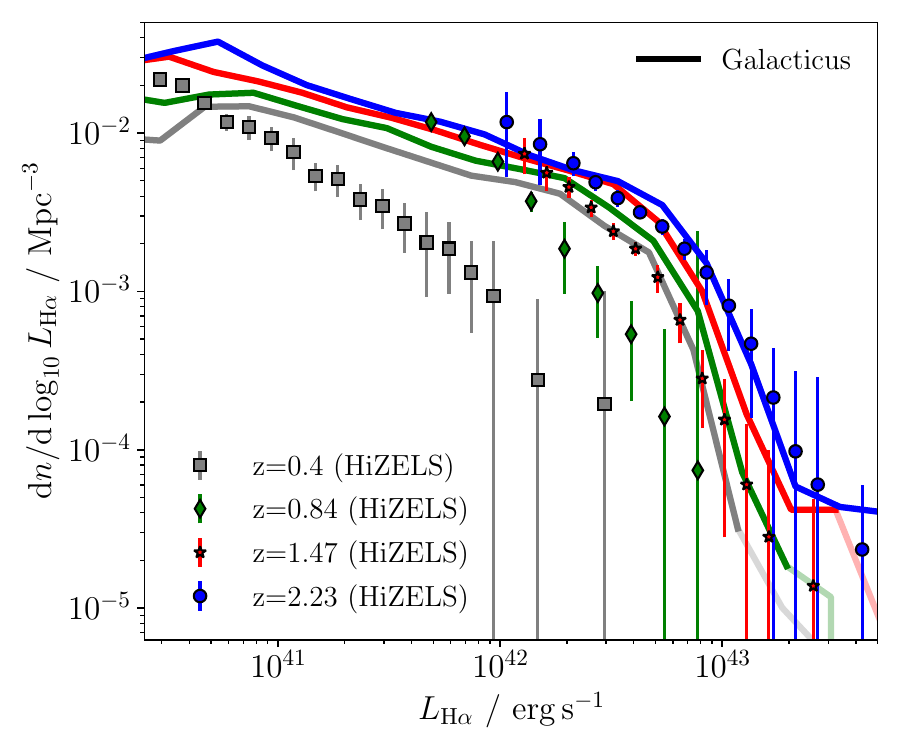} 
	\caption{H$\alpha$ luminosity functions from our calibrated \galacticus model (Section~\ref{sect:includingAdditionalConstraints}) compared with measurements from the HiZELS survey \citep{2013MNRAS.428.1128S}. The HiZELS luminosities shown have been corrected for dust attenuation by applying a uniform 1\,mag brightening to all emission lines. To enable a direct comparison, we therefore plot the \emph{dust-free} \galacticus predictions, with colours indicating redshift to match the observational points. The \galacticus lines become semi-transparent for $L_{\mathrm{H}\alpha}$-bins containing fewer than 10 galaxies. The model reproduces the observations well at $z=2.23$, but shows little redshift evolution and severely over predicts the number of H$\alpha$-bright galaxies by $z=0.4$.}
	\label{fig:Sobral13comparison}
\end{figure}

In Fig.~\ref{fig:Sobral13comparison} we compare the $\mathrm{H}\alpha$ luminosity functions predicted by our calibrated \galacticus model with dust-corrected measurements from the HiZELS survey \citep{2013MNRAS.428.1128S}. At the highest redshift probed, $z=2.23$, the model reproduces both the normalisation and the shape of the observed luminosity function across the full luminosity range.  

At $z=0.4$, however, the model substantially over-predicts the number density of $\mathrm{H}\alpha$-bright galaxies. Dust attenuation of nebular lines is observed to be somewhat stronger at low redshift \citep[e.g.][]{2013ApJ...763..145D}, so a constant dust correction, as applied by \citet{2013MNRAS.428.1128S}, may indeed under-correct the low-$z$ luminosity function relative to higher $z$. However, the measured redshift dependence is too weak to explain the excess we find. In addition, the slope of the model luminosity function at $z=0.4$ is too shallow, meaning that reconciling the model with the data via dust alone would require attenuation to increase steeply with $L_{\mathrm{H}\alpha}$. Observationally, nebular attenuation does correlate with intrinsic line strength: using a large SDSS sample, \citet{2012MNRAS.421..486X} found $E(B-V)$ to rise with H$\alpha$ luminosity. Yet, their results imply only $\sim0.25$\,mag more attenuation from $L_{\mathrm{H}\alpha}=10^{41}$ to $10^{42}\,\mathrm{erg\,s^{-1}}$, whereas steepening our $z=0.4$ H$\alpha$ luminosity function to match the observations would require a $\gtrsim 1$\,mag increase over this range.  

The poor agreement at low redshift therefore cannot be explained solely by dust attenuation, and more likely reflects an overproduction of high star formation rate galaxies in \galacticus at late times. This interpretation is consistent with the evolution of the SHMR shown in Fig.~\ref{fig:SHMR_lowz_and_highz}, where stellar masses at the high-mass end are too low at $z \sim 0.9$ but match observations by $z \sim 0.3$, implying star formation rates that are too high over this interval. While the calibrated model captures the broad behaviour of the H$\alpha$ luminosity function at $z \gtrsim 1$, understanding the origin of the excess of highly star-forming galaxies at low redshift requires a more direct examination of the star formation rates predicted by the model. We turn to this in the next section by analysing the joint distribution of stellar mass and star formation rate and comparing it to empirical determinations of the star-forming main sequence.

\subsection{Star formation rates}
\label{sec:SFRs}

\begin{figure*}
        \centering
        \includegraphics[width=0.99\textwidth]{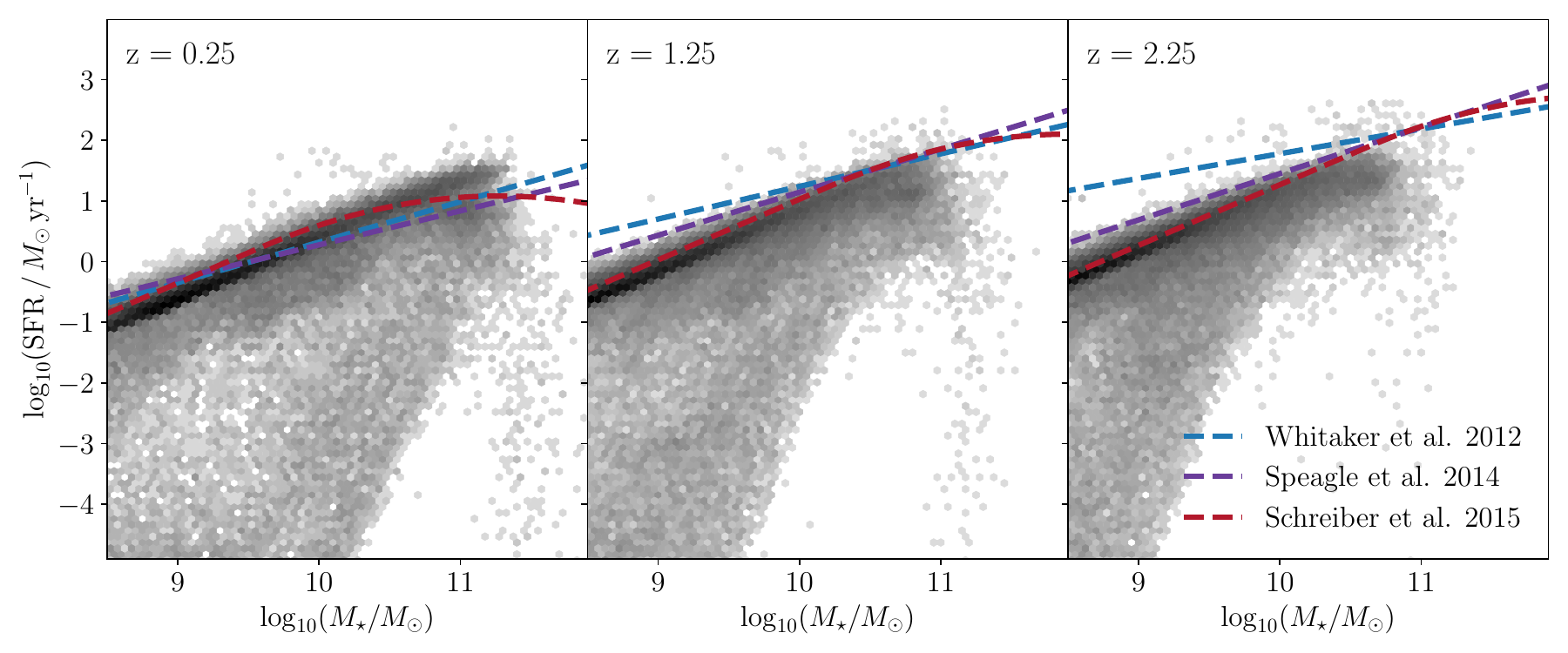} 
        \caption{Stellar mass -- star formation rate relation for \galacticus galaxies at $z=0.25$, $1.25$, and $2.25$. Each panel shows the number density of galaxies in the $\log_{10} M_\star–\log_{10} \mathrm{SFR}$ plane using hexagonal binning, with a logarithmic colour scale. The dashed lines show literature determinations of the star-forming main sequence at the corresponding redshift from \citet{2012ApJ...754L..29W}, \citet{2014ApJS..214...15S} and \citet{2015A&A...575A..74S}.}
	\label{fig:Mstar_SFR}
\end{figure*}

Fig.~\ref{fig:Mstar_SFR} compares the joint distribution of stellar mass and star formation rate predicted by \galacticus to several commonly used empirical parameterisations of the star-forming main sequence: equation~1 from \citet{2012ApJ...754L..29W}; equation~28 from \citet{2014ApJS..214...15S}; and equation~9 from \citet{2015A&A...575A..74S}. At all three redshifts, the model produces a well-defined locus of star-forming galaxies -- the ``main sequence'' (MS) -- whose normalisation and slope evolve with redshift in broad qualitative agreement with these literature relations. While the \galacticus MS differs from the literature ones in detail, the differences are comparable to the differences between the different observational determinations.\footnote{Observed main-sequence relations can differ, for example, because star-formation rates are inferred using different indicators that probe star formation over different timescales and are affected differently by dust, resulting in systematic offsets between studies.}

The low-redshift panel ($z=0.25$) is particularly informative. In this regime, the \galacticus MS lies at approximately constant specific star formation rate, whereas empirical MS fits tend to be somewhat flatter than this. As a result, the model predicts an excess of high-SFR systems at $M_\star \sim 10^{11} \msun$. This behaviour provides a natural explanation for the over-prediction of bright objects in the low-redshift H$\alpha$ luminosity function seen in Fig.~\ref{fig:Sobral13comparison}, since H$\alpha$ luminosity approximately traces the instantaneous star formation rate. At higher redshift, the differences between the model and the various empirical main-sequence relations are reduced, and the overall distribution of star formation rates at fixed stellar mass is more consistent with observational expectations.

While not a primary focus of this work -- owing to a calibration strategy that targets central galaxies only -- we inspected the star formation properties of satellite galaxies in the calibrated model. In particular, we examined the quenched fraction of satellites and found poor agreement with observational measurements. The model predicts quenched fractions of $\gtrsim 80\%$ for satellite galaxies at all stellar masses, in contrast to observations that show a strong stellar mass dependence, increasing from $\sim 30$--$60\%$ at $M_\star \sim 10^{10}\,\msun$ to $\sim 80$--$90\%$ at $M_\star \sim 10^{11}\,\msun$ \citep{2012MNRAS.424..232W}. This limitation should be borne in mind when interpreting discrepancies involving emission-line statistics that receive contributions from satellite galaxies.

\subsection{Galaxy sizes}
\label{sec:sizes}

We compare the predicted stellar mass--size relation from our fully calibrated model to the 3D-HST+CANDELS measurements of \citet{2014ApJ...788...28V} in three redshift slices ($z\simeq0.25$, $1.25$, and $2.25$), shown in Fig.~\ref{fig:sizes_through_time}. In \citet{2014ApJ...788...28V}, galaxies were determined to be star-forming or quiescent based on cuts in rest-frame $U-V$ and $V-J$ colours. We found, as also seen in \citet{2015MNRAS.451.2663H}, that our galaxies had a bimodal distribution in this colour-colour space, but that the cuts used by \citet{2014ApJ...788...28V} would not neatly separate these two loci -- presumably due to short-comings in the stellar population synthesis or in the modelling of dust effects on the colours. Rather than defining our own colour cuts, we used a physically motivated definition of star-forming and quiescent, classifying galaxies based on their offset from the star-forming main sequence,
\[
\Delta_{\rm MS}\equiv \log_{10}{\rm SFR}-\log_{10}{\rm SFR}_{\rm MS}(M_\star,z),
\]
where ${\rm SFR}_{\rm MS}(M_\star,z)$ is taken from \citet{2015A&A...575A..74S}, owing to it most faithfully matching the main sequence of our \galacticus model (Fig.~\ref{fig:Mstar_SFR}). We label galaxies as quiescent if $\Delta_{\rm MS}\le -1$\,dex and star-forming otherwise. For each class we compute the median and 16th--84th percentile range of the 3D half-light radii $r_{50}$ in bins of stellar mass and compare these to the observational medians and percentile ranges.

\subsubsection{Size definitions and comparability}
\label{sect:size_definitions}

In our model, $r_{50}$ denotes a three-dimensional (spherical) half-mass radius for stars, as determined by \galacticus. In the observations of \citet{2014ApJ...788...28V}, the reported size is the \emph{semi-major-axis effective radius}, i.e. the semi-major axis of the ellipse that encloses half of the total flux of the best-fitting S\'ersic model. These definitions are not identical, which can introduce systematic offsets.

For an axisymmetric, infinitesimally thin disk, the semi-major-axis effective radius is inclination-independent and equal to the intrinsic in-plane half-light radius of the disk, which (owing to the mass all lying in-plane) is also equal to the 3D half-light radius. 
For a spherically symmetric Hernquist profile (which is the spheroid model we use in \galacticus) the three-dimensional and projected half-light radii differ by a fixed geometric factor: $r_{50}^\mathrm{3D} \approx 1.33 \, R_{50}^\mathrm{2D}$ \citep{1990ApJ...356..359H}.  

In practice, real galaxies depart from the idealised morphologies assumed by \galacticus: disks can exhibit non-axisymmetric structure (e.g.\ bars and spiral arms), spheroids can be triaxial, and dust plus mass-to-light gradients affect light-weighted sizes. These effects can alter the mapping between three-dimensional and projected sizes relative to the simple cases discussed above.

The relationship between 3D and 2D size measurements has been studied using hydrodynamical simulations. For example, \citet{2017MNRAS.465..722F} report that the semi-major-axis effective radii \citep[as used by][]{2014ApJ...788...28V} are typically smaller than three-dimensional half-light sizes by $0.1-0.2$ dex, with little dependence on stellar mass and redshift. We therefore interpret our size comparisons with these systematic differences in mind.

\subsubsection{Comparison with the data}

Across the three redshift slices, the model reproduces the overall growth of galaxy sizes with cosmic time at fixed $M_\star$. For star-forming systems, the median $r_{50}$ generally lies within 0.2\,dex of the observed medians with comparable scatter, though the mass--size relation is somewhat too shallow: high-mass galaxies ($M_\star\gtrsim10^{10.7}\,\mathrm{M_\odot}$) tend to be smaller than observed, while low-mass galaxies are larger. Quiescent galaxies are systematically too large at all redshifts, and the high-mass slope likewise appears too flat. At the highest masses the model shows a turnover (decreasing $r_{50}$) for both star forming and quiescent galaxies, which is not seen in the data. These trends are qualitatively consistent across bins and are robust to modest changes in the quiescent selection threshold (e.g. $\Delta_{\rm MS}=-0.8$ to $-1.2$\,dex), and to using a different definition of the main sequence \citep[e.g. using][]{2014ApJS..214...15S}.

\begin{figure*}
        \centering
        \includegraphics[width=0.99\textwidth]{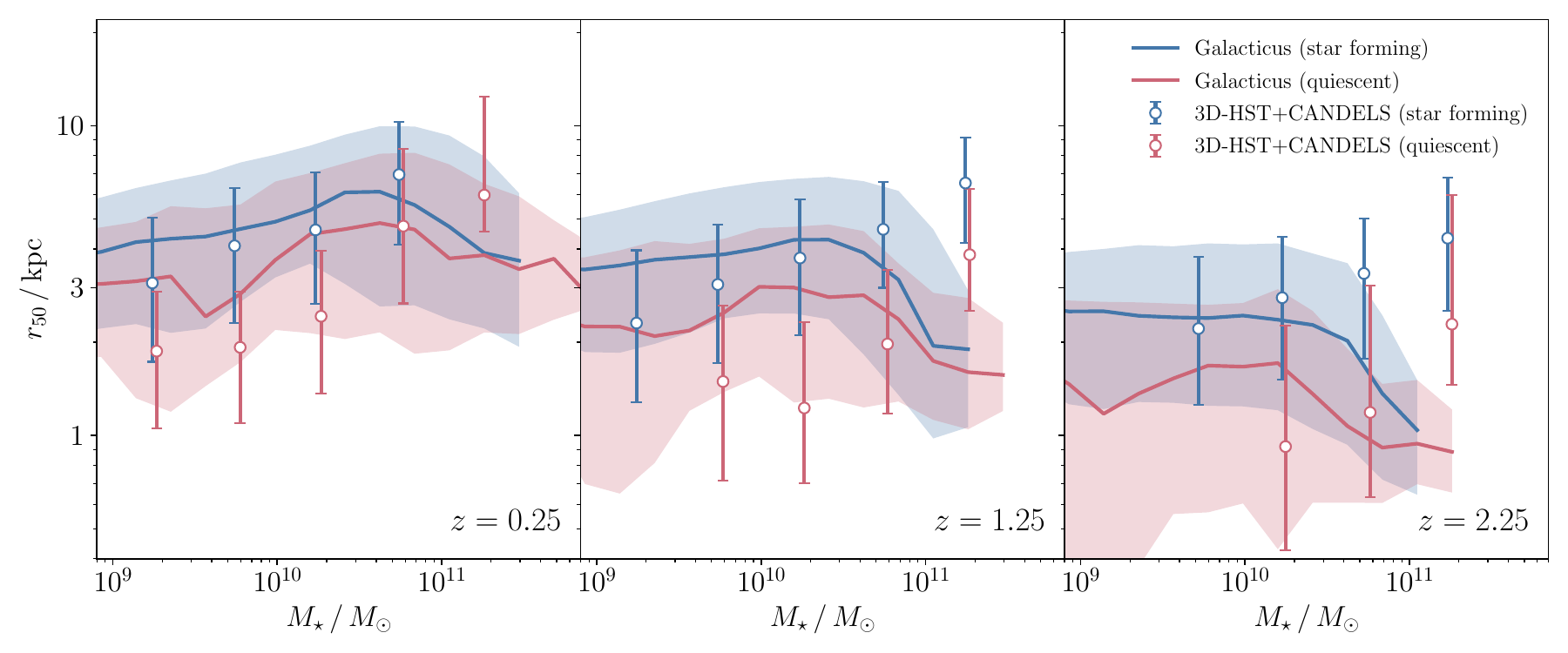} 
        \caption{Stellar mass–size relation at $z=0.25$, $1.25$, and $2.25$. Each panel shows the half-light radius $r_{50}$ as a function of stellar mass $M_\star$ for star-forming (blue) and quiescent (red) galaxies. Shaded bands are the 16th–84th percentile ranges of the \galacticus predictions; solid curves are the corresponding medians. Points with error bars give the 3D-HST+CANDELS medians and 16th–84th percentile ranges from \citet{2014ApJ...788...28V}. The \galacticus results for each galaxy type cover the stellar mass range in which there are $\geq 10$ galaxies per bin. The model reproduces the growth of galaxy sizes with cosmic time at fixed $M_\star$, but its mass–size relation is shallower than observed and shows a turnover (decreasing $r_{50}$) at the highest stellar masses.}

	\label{fig:sizes_through_time}
\end{figure*}

\section{Discussion}
\label{sect:discussion}

\subsection{Matching Across Redshift}
\label{sect:matchAcrossRedshift}

Our accelerated calibration strategy succeeds in reproducing several key observables, but also highlights both methodological and physical challenges that remain. On the positive side, by calibrating \galacticus to the SHMR in just a few halo mass bins, we recover a low-redshift stellar mass function that matches observations to high accuracy. However, important discrepancies persist. The most notable is the inability of our calibrated model to reconcile the SHMR at both low and high redshift simultaneously: the relation is too flat at $z \sim 1$, producing an excess of low-mass galaxies and too few high-mass systems.

This tension reflects a long-standing difficulty in semi-analytic models, seen also in earlier work \citep[e.g][]{2013MNRAS.428.1351G, 2013MNRAS.431.3373H}. Our adoption of the \citet{2013MNRAS.431.3373H} gas recycling timescale model was intended to provide additional freedom to address this, but the recycling parameters are only weakly constrained by our calibration, with posterior distributions that remain broad. Tests in which we reduced recycling times by an order of magnitude produced little change in the predicted galaxy properties, suggesting that cooling of the hot halo is sufficiently slow that the timescale for recycling makes little difference to the supply of gas for star formation. Future work may therefore require either more radical changes to the gas cooling and recycling models or alternative parameterisations of feedback.

\subsection{Galaxy Sizes}
\label{sect:improveSizesMatch}

Galaxy sizes present another limitation of our model. While we reproduce the overall increase of galaxy sizes with cosmic time, the predicted mass--size relation is too shallow: low-mass galaxies are too large, while high-mass galaxies are too small. Our calibration dataset included the low-$z$ mass--size relation, but the \galacticus model parameter space explored by our MCMCs offered limited freedom to adjust galaxy sizes.  

Galaxy sizes depend on the baryon cycle in a non-trivial way. The specific angular momentum of the gas reflects that of the halo at the time of cooling, and because the distribution of dimensionless halo spin parameters is roughly constant through time, while halos grow in mass and size, the angular momentum of accreted gas increases with cosmic time. This imprint of halo growth on the cooling history contributes to the redshift evolution of galaxy sizes.  

In our calibration, the only parameter directly associated with galaxy sizes that we varied was $R_2/R_3$. We freed this parameter from its default value because early tests (not shown here) produced spheroid sizes that were considerably too large. Although spheroid sizes can indirectly influence other galaxy properties (e.g. star formation efficiency, disk growth), the explicit calibration of $R_2/R_3$ relied on the bulge-dominated systems in our set of 27 low-$z$ halos. As shown in the $M_\star$--$R_\star^{50}$ panel of Fig.~\ref{fig:fullCalibrationMCMC}, only four of these galaxies are bulge-dominated, all with similar stellar masses. Consequently, the calibration of spheroid sizes was essentially unconstrained with respect to the stellar mass dependence seen in observations.  

A more effective approach might be to calibrate simultaneously to the bulge and disk sizes of all galaxies, rather than dividing the population into bulge- or disk-dominated systems. This would allow the sizes of spheroids in disk-dominated galaxies, as well as the sizes of disks in bulge-dominated galaxies, to inform the fit. Suitable data for such an approach exist, for example in \citet{2006MNRAS.368..414D} and \citet{ 2016MNRAS.462.1470L}, and could provide the necessary leverage to more robustly constrain size-related parameters, to improve the stellar mass dependence of galaxy sizes in future calibrations.


\subsection{Future Directions}

Looking forward, the main challenge is to introduce sufficient physical flexibility into \galacticus to enable a simultaneous match to multiple datasets across redshift. There are several possible avenues that could help to address the tensions we have identified. For example, greater freedom in the treatment of gas cooling and recycling, alternative prescriptions for stellar and AGN feedback, or a more flexible modelling of galaxy sizes could all plausibly improve agreement with observations. At present it is not clear which of these modifications is most critical -- or whether changes in just one area might be sufficient -- but they provide natural starting points for future investigation.  

The calibration framework we have presented provides a flexible platform for such developments. Extending it to additional observables is not entirely straightforward, however. Our method relies on targeting specific halo masses, while most galaxy properties are measured across the full galaxy population rather than at fixed halo mass. One possible route forward would be to use results from empirical models \citep[such as the \textsc{UniverseMachine},][]{2019MNRAS.488.3143B}, which provide relations between halo mass and galaxy properties such as star formation rates. These relations could then be used in a similar way to the SHMR in this work, allowing us to target particular halo masses and ask whether the predicted galaxy populations match what is inferred from an empirical model calibrated on real observations. In this way, constraints from multiple galaxy properties, sensitive to different aspects of galaxy evolution, could be brought together in a consistent calibration framework. Furthermore, when combined with more efficient optimisation algorithms (e.g. particle swarm optimisation), this approach could explore a wide variety of parameterisations and physical prescriptions at relatively low computational cost.  

A natural workflow would be to adopt a two-stage strategy. First, one could explore a broad range of physical prescriptions and combinations of free parameters using a fast likelihood of the type introduced here. This stage would identify a small number of promising models or parameterisations. In the second stage, the most promising models could then be studied in detail by thoroughly exploring their parameter space and calibrating directly against observed quantities using emulation.  

Modern emulation techniques (e.g. Gaussian processes and neural networks) have already been applied to semi-analytic models \citep{2010MNRAS.407.2017B,2014arXiv1405.4976V,2021MNRAS.506.4011E}. To achieve a suitable level of predictive accuracy, one might expect to require of order $\sim 1000$ model runs spread across the parameter space. Each of these runs would simulate large numbers of halos, enabling robust calculation of observables such as luminosity functions, stellar mass functions, atomic gas mass functions, etc. For any particular observable, an emulator can then be trained on this simulated dataset. Once trained, the emulator provides near-instantaneous predictions of the observable for arbitrary parameter combinations. This dramatically reduces the cost of mapping the parameter space, making it feasible to apply Bayesian inference methods such as MCMC to quantify parameter constraints and degeneracies.  

In summary, our accelerated calibration approach should be seen as complementary to emulation. The former provides a rapid means of sifting through large model spaces to identify promising prescriptions, while the latter allows detailed posterior inference once a model family has been chosen. Combining the two approaches offers a clear path towards robust, flexible, and computationally efficient calibration of semi-analytic galaxy formation models in preparation for upcoming large-scale structure surveys.

\section{Conclusions}
\label{sect:conclusions}

We have presented an accelerated calibration framework for semi-analytic galaxy formation models, demonstrated here with \galacticus. The central idea is to construct a fast likelihood based on the stellar-to-halo mass relation (SHMR) evaluated at a small number of target halo masses, so that each likelihood evaluation requires simulating only tens of galaxies. This renders MCMC-based calibration practical for \galacticus, which would be prohibitively expensive to calibrate using full-population observables.

Calibrating to the low-redshift SHMR, we recover a realistic low-redshift stellar mass function (Fig.~\ref{fig:SHMRcalibration}). We then extend the fast likelihood to include additional constraints (specifically a higher-redshift SHMR and the local stellar mass--size relation), demonstrating that the same machinery can incorporate qualitatively different datasets.

This exercise exposes important limitations that guide future development. The calibrated model does not simultaneously reproduce the SHMR at both low and high redshift: at $z\sim1$ the relation is overly shallow (Fig.~\ref{fig:SHMR_lowz_and_highz}). The mass--size relation is also too flat, with disk sizes typically too large and bulge sizes showing little stellar-mass dependence (Fig.~\ref{fig:MstarR50}). These results indicate that additional physical freedom is required for a simultaneous match to these datasets. Crucially, the framework introduced here is well-suited to exploring such alternative parameterisations systematically and at low computational cost.

The broader diagnostics in Section~\ref{sect:results} show that the stellar mass function remains generally good at low redshift but increasingly over-predicts the abundance of low-mass galaxies toward higher redshift (Fig.~\ref{fig:ZFOURGEcomparison}), in line with the too-shallow high-$z$ SHMR. The H$\alpha$ luminosity function is well reproduced at $z \approx 2.2$, whereas by $z \approx 0.4$ the model yields too many H$\alpha$-bright systems (Fig.~\ref{fig:Sobral13comparison}), indicating excessive late-time star formation. By analysing the joint $M_\star - \mathrm{SFR}$ distribution of our model, we see that this excess star formation happens primarily in massive galaxies (Fig.~\ref{fig:Mstar_SFR}). Size evolution with redshift is captured in broad terms: median sizes grow with cosmic time for both star-forming and quiescent systems (Fig.~\ref{fig:sizes_through_time}). However, at fixed redshift the stellar mass--size relation is too shallow, with star-forming galaxies too large at low masses and quiescent galaxy sizes showing almost no stellar-mass dependence, in contrast with what is observed.

Looking ahead, we envision a two-stage strategy. First, use the fast, SHMR-based calibration to scan model choices and parameterisations efficiently, identifying a small set of promising variants. Second, apply emulator-based inference to those variants, enabling joint fits to population-wide observables (e.g.\ stellar mass functions, luminosity functions, and size distributions). 
With upcoming surveys from Euclid, Roman and Rubin set to deliver vast, high-quality datasets, the ability to iterate quickly on SAM physics will be essential. The present work demonstrates a practical pathway toward that goal, with a calibration workflow that is rapid enough to support extensive experimentation.

\section*{Data Availability}

The \galacticus\ configuration files used to run our MCMCs, the resulting chains, and the scripts that generate all figures are available at \href{https://github.com/Andrew-Robertson/CalibratingGalacticus-Paper1}{github.com/Andrew-Robertson/CalibratingGalacticus-Paper1}. 
The \galacticus\ galaxy catalogues associated with this work (e.g. those used in Section~\ref{sect:results}) are deposited on Zenodo at \href{https://doi.org/10.5281/zenodo.16952803}{doi.org/10.5281/zenodo.16952803}. 
These catalogues contain metadata describing the exact \galacticus\ version used, as well as all parameter values.

\section*{Acknowledgments}

AR gratefully acknowledges funding from NASA Grant  \#80NSSC24M0021, ``Project Infrastructure for the Roman Galaxy Redshift Survey''. Calculations in this work were performed using the OBS HPC resource operated by Carnegie Science. This work made use of the following software packages: \href{https://github.com/astropy/astropy}{{Astropy}}
\citep{astropy1, astropy2},
\href{https://getdist.readthedocs.io/en/latest/intro.html}{{GetDist}}
\citep{2019arXiv191013970L},
\href{https://github.com/matplotlib/matplotlib}{{Matplotlib}}
\citep{matplotlib},
\href{https://github.com/numpy/numpy}{{NumPy}}
\citep{numpy},
and
\href{https://github.com/scipy/scipy}{{Scipy}}
\citep{scipy}.




\bibliographystyle{aasjournal}
\bibliography{bibliography}


\appendix
\section{Quasi-likelihood for the stellar mass--size relation}
\label{app:size_quasilike}

In Section~\ref{sect:mass_size_likelihood} we presented the likelihood that we use to compare \galacticus\ sizes with those measured by \citet{2003MNRAS.343..978S}, which we copy here for convenience:
\begin{equation}
\mathcal{L}_\mathrm{size}
= \prod_{b=1}^{3}\;\prod_{i=1}^{N_b}
p_{\mathrm{obs}}^{(t_{bi})}\!\big(R_{\star,bi}^{50}\mid M_{\star,bi}\big),
\end{equation}
where $t_{bi}\in\{\mathrm{early},\mathrm{late}\}$ is the morphological type assigned via $B/T$.

A few comments on this method are in order. Firstly, it will clearly tend to drive the \galacticus\ parameters towards those that produce sizes close to the median $M_\star$--$R_\star^{50}$ relations from \citet{2003MNRAS.343..978S}, which is the purpose of this term in the likelihood. That said, it is not formally correct. Perhaps the easiest way to see this is that the likelihood is not maximised by a set of sizes that are distributed according to the \citet{2003MNRAS.343..978S} $M_\star$--$R_\star^{50}$ relation with its scatter, but would instead be maximised by every galaxy lying on the median $M_\star$--$R_\star^{50}$ relation from \citet{2003MNRAS.343..978S} (with zero scatter). In this zero-scatter limit, \galacticus\ would produce a distribution of sizes that is inconsistent with the observed Universe, but would nevertheless be preferred by our likelihood.

This may seem counter-intuitive, given that likelihoods which are products of probability densities are ubiquitous when fitting models to data. What is distinct about our case is that the ``data'' here is an observationally inferred \emph{probability distribution} for galaxy sizes, while the ``model'' provides a set of galaxy sizes; our likelihood then evaluates the probability density of drawing those model sizes from the observed distribution. Turning to the $M_\star$--$R_\star^{50}$ panel of Fig.~\ref{fig:fullCalibrationMCMC}, if we imagine (in contrast to what we have here) that the points represented a set of observed galaxies, and that \galacticus\ made a prediction for the distribution of sizes as a function of stellar mass, then \galacticus\ models that predicted too tight a scatter would be penalised by the large $\chi^2$ for galaxies far from the \galacticus\ relation, and models that predicted too broad a scatter would be penalised because a broader log-normal distribution has a lower normalisation. Together, this would favour \galacticus\ models with the correct amount of scatter.

Despite the fact that our method could, in principle, favour \galacticus\ models with a small amount of scatter in size at fixed stellar mass, the maximum-likelihood model shown in Fig.~\ref{fig:fullCalibrationMCMC} indicates that in practice this does not happen. This potentially reflects the fact that galaxies of a given stellar mass have a diversity of assembly histories such that -- at least for our model choices -- it is not easy (or perhaps even possible) for \galacticus\ to produce an extremely tight $M_\star$--$R_\star^{50}$ relation. It could also be that, while a tight $M_\star$--$R_\star^{50}$ relation is achievable, this cannot be done while simultaneously matching the SHMR and its scatter.

\vspace{1ex}

\end{document}